\documentclass[traditabstract]{aa}

\usepackage{graphicx}
\usepackage{txfonts}
\usepackage{natbib}
\usepackage{enumerate}
\usepackage{bm}

\bibpunct{(}{)}{;}{a}{}{,} % to follow A\&A style

\newcommand{\sbs}{SBS\,0335$-$052E}
\newcommand{\cgcg}{CGCG\,007$-$025}
\newcommand{\izw}{I\,Zw\,18}
\newcommand{\iizw}{II\,Zw\,40}

\newcommand{\coone}{$^{12}$CO(1--0)}
\newcommand{\cotwo}{$^{12}$CO(2--1)}

\newcommand{\cothree}{$^{12}$CO(3--2)}

\newcommand{\ta}{T$_{\rm A}^*$}
\newcommand{\tmb}{T$_{\rm mb}$}
\newcommand{\tdep}{$\tau_{\rm dep}$}
\newcommand{\taudust}{$\tau_{\rm dust}$}

\newcommand{\ci}{C{\sc i}}
\newcommand{\cii}{C{\sc ii}}
\newcommand{\ha}{H$\alpha$}

\newcommand{\kms}{km\,s$^{-1}$}
\newcommand{\kkms}{K\,km\,s$^{-1}$}
\newcommand{\htwo}{H$_2$}
\newcommand{\fhtwo}{{\it f}$_{\rm H_2}$}
\newcommand{\gnot}{$G^\prime_0$}
\newcommand{\fco}{{\it f}$_{\rm CO}$}
\newcommand{\av}{A$_V$}

\newcommand{\hst}{{\sl HST}}

\newcommand{\logoh}{12$+$log(O/H)}

\newcommand{\sigmagas}{$\Sigma_{\rm gas}$}
\newcommand{\sigmahi}{$\Sigma_{\rm HI}$}
\newcommand{\sigmahtwo}{$\Sigma_{\rm H2}$}
\newcommand{\mhtwo}{M$_{\rm H2}$}
\newcommand{\mhi}{M$_{\rm HI}$}
\newcommand{\mstar}{M$_{\rm star}$}
\newcommand{\mgas}{M$_{\rm gas}$}
\newcommand{\mbary}{M$_{\rm baryonic}$}
\newcommand{\lco}{L$^\prime_{\rm CO}$}
\newcommand{\aco}{$\alpha_{\rm CO}$}
\newcommand{\sigism}{$\Sigma_{\rm ISM}$}

\newcommand{\cmthree}{cm$^{-3}$}

\newcommand{\hi}{\rm H{\sc i}}
\newcommand{\hii}{\rm H{\sc ii}}

\newcommand{\fev}{\rm [Fe{\sc v}]}
\newcommand{\feii}{\rm [Fe{\sc ii}]}
\newcommand{\oiv}{\rm [O{\sc iv}]}
\newcommand{\oiii}{\rm [O{\sc iii}]}
\newcommand{\micron}{$\mu$m}
\newcommand{\zzsun}{${\mathrm Z/Z}_\odot$}

\newcommand{\subjectfont}{\normalsize\sffamily \selectfont} 
\def\tex {\ifmmode{{T}_{\rm ex}}\else{$T_{\rm ex}$}\fi}
\def\tmb {\ifmmode{{T}_{\rm mb}}\else{$T_{\rm mb}$}\fi}
\def\ci  {\ifmmode{{\rm C}{\rm \small I}}\else{C\ts {\scriptsize I}}\fi}
\def\hi  {\ifmmode{{\rm H}{\rm \small I}}\else{H\ts {\scriptsize I}}\fi}
\def\hh  {\ifmmode{{\rm H}_2}\else{H$_2$}\fi}
\def\kms    {\ifmmode{{\rm \ts km\ts s}^{-1}}\else{\ts km\ts s$^{-1}$}\fi}
\def\msun   {\ifmmode{{\rm M}_{\odot}}\else{M$_{\odot}$}\fi}
\def\msunpc   {\ifmmode{{\rm M}_{\odot}\,{\rm pc}^{-2}}\else{M$_{\odot}$\,pc$^{-2}$}\fi}
\def\msunyr   {\ifmmode{{\rm M}_{\odot}\,{\rm yr}^{-1}}\else{M$_{\odot}$\,yr$^{-1}$}\fi}
\def\lsun   {\ifmmode{{\rm L}_{\odot}}\else{L$_{\odot}$}\fi}
\def\zsun   {\ifmmode{{\rm Z}_{\odot}}\else{Z$_{\odot}$}\fi}

\begin{document}

\title{Molecular depletion times and the CO-to-\htwo\ conversion factor in metal-poor galaxies
\thanks{Based on observations carried out with the IRAM 30m; 
IRAM is supported by the INSU/CNRS (France), MPG (Germany), and IGN (Spain).
}
}

\author{L.~K. Hunt \inst{\ref{inst:hunt}}
\and
S. Garc\'{\i}a-Burillo \inst{\ref{inst:garciaburillo}}
\and
V. Casasola \inst{\ref{inst:hunt}}
\and
P. Caselli \inst{\ref{inst:caselli}}
\and
F. Combes \inst{\ref{inst:combes}}
\and
C. Henkel \inst{\ref{inst:henkela},\ref{inst:henkelb}}
\and
A. Lundgren \inst{\ref{inst:lundgren}}
\and
R. Maiolino \inst{\ref{inst:maiolino}}
\and
K.~M. Menten \inst{\ref{inst:henkela}}
\and
L. Testi \inst{\ref{inst:testi}}
\and
A. Weiss \inst{\ref{inst:henkela}}
}

\offprints{L. K. Hunt}
\institute{INAF - Osservatorio Astrofisico di Arcetri, Largo E. Fermi, 5, 50125, Firenze, Italy
\label{inst:hunt}
\email{hunt@arcetri.astro.it}
 \and
Observatorio Astron\'omico Nacional (OAN)-Observatorio de Madrid,
Alfonso XII, 3, 28014-Madrid, Spain
\label{inst:garciaburillo}
 \and
Max-Planck-Institut f\"ur extraterrestrische Physik, Giessenbachstrasse 1, 85748 Garching, Germany
\label{inst:caselli}
 \and
Observatoire de Paris, LERMA, College de France, CNRS, PSL, Sorbonne University UPMC, F-75014, Paris, France
\label{inst:combes}
 \and
Max-Planck-Institut f\"ur Radioastronomie, Auf dem H\"ugel 69, 53121 Bonn, Germany
\label{inst:henkela}
 \and
Astronomy Department, King Abdulaziz University, P.O. Box 80203, Jeddah, Saudia Arabia
\label{inst:henkelb}
 \and
ALMA JAO, Alonso de Cordova 3107, Vitacura, Casilla 19001, Santiago, Chile
\label{inst:lundgren}
 \and
Cavendish Laboratory, University of Cambridge, 19 J.J. Thomson Avenue, Cambridge CB3 0HE, UK
\label{inst:maiolino}
\and
ESO, Karl Schwarzschild str. 2, 85748 Garching bei M\"unchen, Germany
\label{inst:testi}
}

   \date{Received  2015/ Accepted  2015}

   \titlerunning{Molecular depletion times at low metallicity}
   \authorrunning{Hunt et al.}

\abstract{Tracing molecular hydrogen content with carbon monoxide in low-metallicity galaxies 
has been exceedingly difficult.
Here we present a new effort, with IRAM 30-m observations of \coone\ of a sample of 8 
dwarf galaxies having oxygen abundances ranging from \logoh$\sim$7.7 to 8.4.
CO emission is detected in all galaxies, including the most metal-poor galaxy of our sample
(0.1\,\zsun); to our knowledge this is the largest number of
\coone\ detections ever reported for galaxies with \logoh$\la$8 (0.2\,\zsun) outside the Local Group.
We calculate stellar masses, \mstar, and star-formation rates (SFRs), and analyze
our results in conjunction with galaxy samples from the literature.
Extending previous results for a correlation of the molecular gas depletion time,
\tdep, with \mstar\ and specific SFR (sSFR), we find a variation in \tdep\
of a factor of 200 or more (from $\la$50\,Myr to $\sim$10\,Gyr) over a spread
of $\sim 10^3$ in sSFR and \mstar.
We exploit the variation of \tdep\ to constrain the CO-to-\htwo\ mass conversion
factor \aco\ at low metallicity, and assuming a power-law variation find \aco\,$\propto$\,\zzsun$^{-2}$,
similar to results based on dust continuum measurements compared with gas mass.
By including \hi\ measurements, we show that the fraction of total gas mass relative to the baryonic
mass is higher in galaxies that are metal poor, of low mass, and with high sSFR. 
Finally, comparisons of the data with star-formation models of the molecular gas phases show that
the models are generally quite successful, but
at metallicities \zzsun$\la$0.2, there are some discrepancies. 

}
\maketitle

%---------------------------------------------------------------

\section{Introduction}
\label{sec:intro}

Dense clumps in cool molecular clouds are thought to be the ``cradles"
of star formation, but finding molecules at low metallicities 
has been exceedingly difficult.
In dwarf galaxies with nebular oxygen abundances 
$\la0.3$\,\zsun\ (\logoh$\la$8.2)\footnote{Throughout this work, 
we will rely on the \citet{asplund09} solar abundance calibration of \logoh\,=\,8.69.}, 
despite vigorous
on-going star formation, there seems to be very little CO
\citep{sage92,taylor98,gondhalekar98,barone00,leroy05,buyle06,leroy07,schruba12,cormier14}.
Intriguingly, this is the same metallicity ``threshold'' below
which little or no polycyclic aromatic hydrocarbon (PAH) emission
is detected in metal-poor galaxies 
\citep{engelbracht05,wu06,madden06,hunt10}.

The reason for this deficit in CO at low metallicities is not yet completely
understood, although there is strong evidence for ``CO-dark'' molecular gas
in the interstellar medium (ISM) \citep[e.g.,][]{wolfire10,leroy11,planck11,paradis12,lebout12,pineda14}.
Such a component would result from a diffuse or metal-poor ISM in which
CO suffers from enhanced photodissociation relative to \htwo\ \citep[e.g.,][]{maloney88,bolatto99,glover11,narayanan12}. 
\htwo\ can effectively avoid the photodissociation effects of UV radiation
through self-shielding, even at relatively low gas surface densities, \sigism.
On the other hand, CO needs higher \sigism\ to self-shield, thus 
implying that the CO-to-\htwo\ ratio should vary even within an individual molecular
cloud \citep[e.g.,][]{rollig06,leroy13a,hughes13}.
Under conditions of low \sigism\ and low dust extinction \av,
most of the carbon would be in other forms such as \ci\ or \cii,
rather than in CO.

Nevertheless,
even under extreme conditions in a metal-poor ISM, it is likely that stars form
in cool, dense molecular clouds \citep[e.g.,][]{krumholz11}.
However, \htwo, the dominant molecular component of giant molecular clouds, is not directly observable,
but rather inferred from
line emission of rotational transitions of \coone, unfortunately deficient at low metallicity.
Constraining the properties of metal-poor molecular clouds observationally is difficult 
because the factor, \aco, relating
CO luminosity \lco\ to \htwo\ mass \mhtwo\ can vary dramatically, because of the
presence of CO-dark gas.
\aco\ can be more than
%10 times lower than the Galactic value in dense, molecular
%starbursts and in the central regions of some disk galaxies including
%our own 
%\citep{blitz85,sodroski95,scoville97,tacconi08,sandstrom13},
50 times higher than the Galactic value in metal-poor
environments and in regions of low \av\ and low \sigism\ 
\citep{israel97,leroy09b,leroy11,pineda10,bolatto11,schruba12,sandstrom13}.
Although the relation between CO and \htwo\ 
has been the subject of much debate \citep[see][for a recent review]{bolatto13},
% referee
it is clear that 
at some low metallicity, probably $\la0.05-0.1$\,\zsun, CO ceases to become
a useful tracer of total \htwo\ mass.

In this paper, we begin to explore the limits to which CO can trace \htwo\ in
metal-poor star-forming galaxies.
We first describe the targets and ancillary data in Sect.~\ref{sec:sample},
and the \coone\ observations and data analysis in Sect.~\ref{sec:observations}. 
In Sect.~\ref{sec:analysis}, we compare our results with previous work,
and assess the dependence of CO luminosity on star-formation rate (SFR).
Sect.~\ref{sec:depletion} describes
constraints on \aco\ and \tdep, the molecular gas depletion time, i.e., 
the time required to consume the \htwo\ gas reservoir.
We adopt a variation of the method proposed by \citet{saintonge11b}, 
extending it to low metallicity in order to
separate the effects of \tdep\ variations from variations of \aco.
In Sect.\ref{sec:atomic}, we analyze the \htwo\ content of our sample and
compare it with \hi\ for a compilation of galaxies taken from the literature. 
We discuss the implications of our results in Sect.~\ref{sec:discussion},
and, in Sect.~\ref{sec:conclusions},
summarize our results and present our conclusions. 

\section{The targets and the ancillary data\label{sec:sample}}

\begin{figure*}[ht]
\hbox{
\centerline{
\includegraphics[angle=0,width=0.25\linewidth]{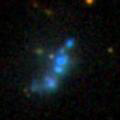}
\includegraphics[angle=0,width=0.25\linewidth]{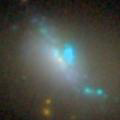}
\includegraphics[angle=0,width=0.25\linewidth]{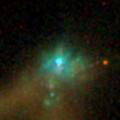}
\includegraphics[angle=0,width=0.25\linewidth]{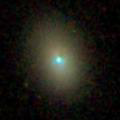}
}
}
\vspace{0.1\baselineskip}
\hbox{
\centerline{
\includegraphics[angle=0,width=0.25\linewidth]{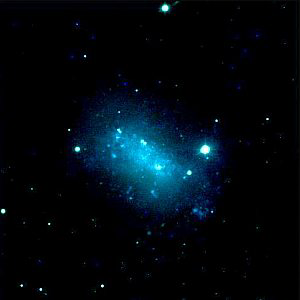}
\includegraphics[angle=0,width=0.25\linewidth]{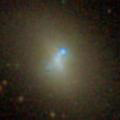}
\includegraphics[angle=0,width=0.25\linewidth]{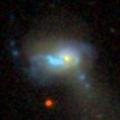}
\includegraphics[angle=0,width=0.25\linewidth]{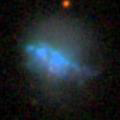}
}
}
\caption{Top row, from left to right in alphabetical order of name: SDSS composite ($ugriz$) images of \cgcg, Haro\,3 (NGC\,3353), \iizw, Mrk\,996.
Bottom row, leftmost panel: NGC\,1156 (image taken from DSS2, reprocessed);
then from left to right SDSS images of NGC\,7077, UM\,448, UM\,462.
Images are oriented with North up, East left, and  
are 50\arcsec\ on a side, except for NGC\,1156 which is $\sim$10\arcmin.
The color of \iizw\ is reddened by $>$2\,mag \av\ from the Galaxy.
NGC\,1140 and \sbs, both of which harbor SSCs and whose images are not from SDSS, 
are shown in Fig. \ref{fig:moreimages}.
}
\label{fig:images}
\end{figure*}

\begin{figure}[!h]
\hbox{
\centerline{
\includegraphics[angle=0,height=0.5\linewidth]{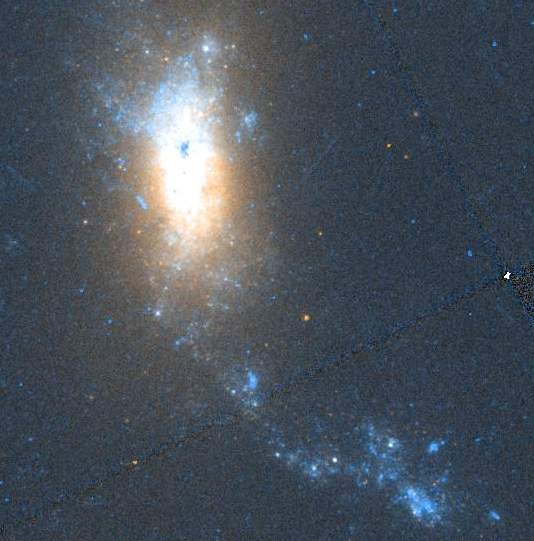}
%\hspace{0.1\baselineskip}
\includegraphics[angle=0,height=0.5\linewidth]{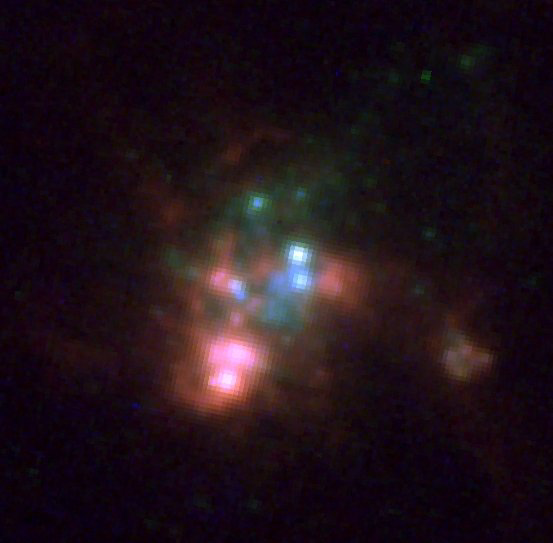}
}
}
\caption{NGC\,1140 and \sbs\ taken from composite \hst\ images;
both galaxies host super-star clusters \citep{hunter94b,reines08}.
The image of NGC\,1140 is $\sim$90\arcsec\ on a side,
and \sbs\ $\sim$5\arcsec, and orientation is North up, East left.
}
\label{fig:moreimages}
\end{figure}

%In this paper we describe only the \coone\ observations, but our main aim in sample selection
%was to observe the higher-J lines, including \cotwo, \cothree, and for selected sources,
%\cofour\ together with isotopologues (\thirteencoone, \thirteencotwo),
%atomic carbon \ci, and dense-gas tracers such as \hcn\
%(these results will be discussed in a companion paper).
%Hence, our sample is mainly equatorial so that we could also observe from the southern
%hemisphere (e.g., APEX).
The observing sample was defined mainly on the basis of detected \htwo\ emission, either
in the ro-vibrational transitions at 2\,\micron, or the purely rotational
transitions in the mid-infrared (e.g., Spitzer/IRS); however, we also targeted
some galaxies that had already been observed in CO
(see Sect. \ref{sec:cocomparison}).
Table \ref{tab:sample} lists the sample galaxies (ordered alphabetically by name), 
together with relevant observational parameters;
Figs. \ref{fig:images} and \ref{fig:moreimages} show thumbprint images. 

% -----------------------------------------------------------------
% --- Table 1: Basic data  
%
\begin{center}
\begin{table*}
      \caption[]{Observing targets} 
\label{tab:sample}
\resizebox{\linewidth}{!}{
%\addtolength{\tabcolsep}{7pt}
{%\small
%\tiny
\begin{tabular}{llrrrcrcrrrc}
\hline
\multicolumn{1}{c}{Name} &
\multicolumn{1}{c}{Other} &
\multicolumn{1}{c}{Redshift$^{\mathrm a}$} &
\multicolumn{1}{c}{Systemic} &
\multicolumn{1}{c}{Distance$^{\mathrm a}$} &
\multicolumn{1}{c}{Distance$^{\mathrm a}$} &
\multicolumn{1}{c}{12$+$} &
\multicolumn{1}{c}{Dimensions$^{\mathrm a}$} &
\multicolumn{1}{c}{SFR$^{\mathrm c}$} &
\multicolumn{1}{c}{Log$^{\mathrm c}$} &
\multicolumn{1}{c}{Log$^{\mathrm d}$} &
\multicolumn{1}{c}{Log(Total)$^{\mathrm e}$} \\
& 
\multicolumn{1}{c}{name} &&
\multicolumn{1}{c}{velocity$^{\mathrm a}$} &
\multicolumn{1}{c}{(Mpc)} &
\multicolumn{1}{c}{method} &
\multicolumn{1}{c}{log(O/H)$^{\mathrm b}$} &
\multicolumn{1}{c}{(arcsec$^2$)} &
\multicolumn{1}{c}{(\msunyr)} &
\multicolumn{1}{c}{\mstar} &
\multicolumn{1}{c}{\mhi} &
\multicolumn{1}{c}{\mhi} \\
& & &
\multicolumn{1}{c}{(\kms)} & & & & & & 
\multicolumn{1}{c}{(\msun)} &
\multicolumn{1}{c}{(\msun)} &
\multicolumn{1}{c}{(\msun)} \\
\multicolumn{1}{c}{(1)} & 
\multicolumn{1}{c}{(2)} & 
\multicolumn{1}{c}{(3)} & 
\multicolumn{1}{c}{(4)} & 
\multicolumn{1}{c}{(5)} & 
\multicolumn{1}{c}{(6)} & 
\multicolumn{1}{c}{(7)} & 
\multicolumn{1}{c}{(8)} & 
\multicolumn{1}{c}{(9)} & 
\multicolumn{1}{c}{(10)} & 
\multicolumn{1}{c}{(11)} & 
\multicolumn{1}{c}{(12)} \\
\hline
\\
CGCG\,007-025   &               &  0.0048 &   1449 & 24.5 & CMB &  7.74 &   27.0 $\times$   16.2  & 0.23  &  8.36  & 8.90 & 8.90 \\ 
\iizw\          & UGCA\,116     &  0.0026 &   789  & 11.7 & CMB &  8.12 &   33.6 $\times$   13.2  & 1.75  &  8.25  & 7.54 & 8.77 \\
Mrk\,996        &               &  0.0054 &   1622 & 18.1 & CMB &  8.00 &   36 $\times$   30      & 0.16  &  9.18  & 8.00 & 8.00 \\
NGC\,1140       & Mrk\,1063     &  0.0050 &   1501 & 19.7 & TF  &  8.18 &  102 $\times$   54    &   0.82  &  9.76  & 8.56 & 9.51 \\
NGC\,1156       &               &  0.0013 &    375 &  8.1 & Stars &  8.23 &  198 $\times$  150  &   0.23  &  9.55  & 8.66 & 9.04 \\
NGC\,3353       & Haro\,3,      &  0.0031 &    944 & 18.1 & TF  &  8.37 &   72.0 $\times$   49.8  & 1.28  &  9.54  & 8.87 & 8.87 \\
                & Mrk\,35 \\
NGC\,7077       & Mrk\,900      &  0.0038 &   1152 & 17.2 & TF  &  8.03 &   48 $\times$   42    &   0.12  &  8.98  & 8.15 & 8.15 \\
\sbs\           &               &  0.0135 &   4043 & 53.6 & CMB &  7.25 &   13.8 $\times$ 12.0  &   1.30  &  8.51  & 7.81 & 8.61 \\
UM\,448         & Arp\,161,     &  0.0186 &   5564 & 81.2 & CMB &  8.04 &   24 $\times$   24    &  11.11  & 10.91  & 9.84 & 9.84 \\
                & UGC\,6665    \\
UM\,462         & Mrk\,1307     &  0.0035 &   1057 & 19.5 & CMB &  7.97 &   36 $\times$   30    &   0.29  &  8.56  & 8.30 & 8.72 \\
                & UGC\,6850   \\ 
\\
\hline
\end{tabular}
}
}
\vspace{0.5\baselineskip}
\begin{description}
\item
[$^{\mathrm{a}}$] NASA/IPAC Extragalactic Database (NED, http://nedwww.ipac.caltech.edu/):
CMB corresponds to Cosmic Microwave Background and TF to Tully-Fisher;
velocities are heliocentric; dimensions are optical major and minor diameters.
\item
[$^{\mathrm{b}}$] Direct-temperature metallicities for 
\cgcg, NGC\,3353, UM\,448, UM\,462, \sbs\ from \citet{izotov07};
\iizw\ \citep{thuan05};
Mrk\,996 \citep{thuan96};
NGC\,1140 \citep{izotov04};
NGC\,1156 \citep{vigroux87};
NGC\,7077 \citep{vanzee06}.
\item
[$^{\mathrm{c}}$]
These quantities are newly derived, as described in Sect. \ref{sec:sfrs}.
\item
[$^{\mathrm{d}}$]
When these values differ from those in Col. (12), they
are calculated from the mean \hi\ surface density integrated
over the optical size of the galaxy.
\item
[$^{\mathrm{e}}$]
The \hi\ measurements are taken from
\citet[][\cgcg, NGC\,7077]{vanzee01},
\citet[][\iizw, UM\,462]{vanzee98},
\citet[][Mrk\,996]{thuan99},
\citet[][NGC\,1140]{hunter94a},
\citet[][NGC\,1156]{swaters02},
\citet[][NGC\,3353]{hunter82},
\citet[][\sbs]{ekta09},
\citet[][UM\,448]{paturel03}.
\end{description}
\end{table*}
\end{center}
% -----------------------------------------------------------------

\subsection{Individual source description \label{sec:source}}

The targets have several characteristics in common, in addition to
their low metallicities.
All of them, with the exception of NGC\,7077 and \sbs,
are sufficiently bright in the infrared (IR) to have been
identified in the (faint) IRAS point-source catalogue.
%(\cgcg, Haro\,3 \iizw, Mrk\,996, NGC\,1140, NGC\,1156, UM\,448, UM\,462)
In fact, \iizw\ was identified as a 60-\micron\ peaker source by \citet{vader93}.
Except for NGC\,7077, all galaxies host conspicuous Wolf-Rayet populations
\citep{schaerer99,guseva00,zhang07};
they are all also rich in atomic gas (see Sect. \ref{sec:atomic}).
This section describes particular features of the individual sources
(in alphabetical order by name as in Table \ref{tab:sample}).

%\subsubsection*{CGCG\,007$-$025}

\medskip
\noindent
{\subjectfont\selectfont $\star$~\cgcg}
%\vspace{-\baselineskip}

\smallskip
\noindent
This galaxy is the most metal-poor (\logoh\,=\,7.7, $\sim$0.1\,\zsun)
\coone\ detection of our sample (see Sect. \ref{sec:analysis}).  
Although it appears in an isolated galaxy catalogue \citep{vanzee01},
it has a companion, a dwarf irregular galaxy at 8.3\,kpc projected
separation (84\,\kms\ velocity separation) which may have triggered
the current star-formation episode \citep{vanzee01}.
It is also found to be a member of a compact group by \citet{focardi02}.
\cgcg\ has the bluest $U-B$ color of any dwarf galaxy in
the \citet{vanzee01} sample, with a very clumpy morphology
as seen in Fig. \ref{fig:images}.
The radiation field is sufficiently hard to excite
\fev, a line that requires ionizing radiation
with photon energies exceeding 4\,Ryd (54.4\,eV)
\citep{thuan05}.
Although a population of Wolf-Rayet stars has been detected in
this galaxy \citep{zhang07},
these are insufficient to power the high-excitation emission lines,
thought instead to be due to fast radiative shocks \citep{thuan05}.
\citet{hunt10} found evidence for rotational \htwo\ emission,
and despite the low metallicity of \cgcg, also 7.7\,\micron\
and 11.3\,\micron\ emission from PAHs.
Judging from the relatively high density of its ionized
gas \citep[$\sim$300\,\cmthree,][]{izotov07}, 
\cgcg\ is an example of an ``active'' starburst,
forming stars in compact dense regions \citep{hirashita04}.

%\subsubsection*{II\,Zw\,40}
\medskip
\noindent
{\subjectfont\selectfont $\star$~\iizw\ (UGCA\,116)}

\smallskip
\noindent
Although we did not observe this source in \coone, it is included
in our CO survey of higher-$J$ lines (Hunt et al., in prep.);
thus we consider it in our sample here.
From its discovery, together with \izw, as an extragalactic \hii\
region \citep{sargent70}, \iizw\ has been the subject of much
scrutiny.
Its two-tailed optical morphology and kinematics suggest that it is the
result of a collision between two gas-rich dwarf galaxies
which triggered the current starburst
\citep{baldwin82,vanzee98}.
Near-infrared recombination lines show that more than half 
the 2\,\micron\ flux comes from free-free emission, and warm
dust contributes at least $\sim$25\% to the 3-4\,\micron\ fluxes
\citep{joy88}.
Emission from the ro-vibrational transitions of \htwo\
at 2\,\micron\ was discovered by \citet{vanzi96}, and later
found to be in a cloud distinct from the primary \hii\ region
\citep{vanzi08}. 
Visual extinction \av\ is spatially variable in \iizw\ and
can be as high as $\ga$10\,mag in the nucleus
\citep{vanzi08}.
This galaxy is undergoing a massive starburst in a
region of $\sim$150\,pc in diameter powered
by $\sim$5000-10000 OV stars \citep{beck02,kepley14}.
These massive stars are driving copious stellar winds which
alter the observed kinematics \citep{bordalo09},
and also produce high-ionization \fev\ \citep{thuan05}.
Like \cgcg, the high density derived from optical spectra
\citep[$\sim500-1400$\,\cmthree,][]{bordalo09} and the small
dimensions of the star-formation episode imply that
it is an ``active'' starburst. 

%\subsubsection*{Mrk\,996}
\medskip
\noindent
{\subjectfont\selectfont $\star$~Mrk\,996}

\smallskip
\noindent
Mrk\,996 is an unusual blue compact dwarf galaxy, with smooth early-type morphology 
and a system of globular clusters \citep{thuan96}.
Virtually all its star formation occurs in the central region
of the galaxy ($\sim$315\,pc in radius).
Because of its high central density ($\sim10^6$\,\cmthree) and
broad emission lines \citep[$\sim$900\,\kms,][]{thuan96},
it was thought that there could be an accreting intermediate-mass
black hole; however X-ray observations revealed no compact
point source \citep{georgakakis11}.
A high ionization \oiv\ line is detected in Mrk\,996
\citep{thuan08}, that, like \fev, needs $\sim$4\,Ryd to excite it. 
Thus, fast radiative shocks are also present in this galaxy,
as its emission-line properties are inconsistent with any other source of
ionization \citep{georgakakis11}.
The lowest rotational transition of \htwo\ ($\sim$28\,\micron) is found in Mrk\,996,
the only one in the sample of \citet{hunt10} in which it was detected.

%\subsubsection*{NGC\,1140}
\medskip
\noindent
{\subjectfont\selectfont $\star$~NGC\,1140 (Mrk\,1063)}

\smallskip
\noindent
NGC\,1140 is a barred dwarf irregular galaxy with peculiar morphology
and a central starburst.
The \ha\ luminosity of its central \hii\ complex is 40 times higher than that
of the giant \hii\ region 30\,Doradus in the Large Magellanic Cloud.
Within this complex, roughly 500\,pc in diameter, there are
6-7 blue super-star clusters (SSCs) \citep{hunter94b,degrijs04}, the
most luminous of these containing almost 6000 O stars \citep{moll07}.
Although there is no galactic-scale outflow in this galaxy,
powerful shocks from the massive SSCs are disrupting the ISM
on scales up to 1-2\,kpc radius \citep{westmoquette10}.
Kinematic and other evidence suggests that the formation of the SSCs 
and the larger-scale starburst have been triggered by an interaction
or merger event with a low-luminosity late-type spiral about 1 Gyr ago 
\citep{hunter94a,hunter94b,westmoquette10}.

%\subsubsection*{NGC\,1156}
\medskip
\noindent
{\subjectfont\selectfont $\star$~NGC\,1156}

\smallskip
\noindent
A typical barred Magellanic irregular about 25\% brighter
in the $B$ band than the Large Magellanic Cloud,
NGC\,1156 is the most quiescent galaxy in our sample.
It has been described as the least disturbed galaxy in the Local
Universe, with no significant galaxy within 700\,kpc
\citep{karachentsev96}. 
More recent \hi\ observations revealed
a single dwarf galaxy at a distance of 80\,kpc
\citep{minchin10} which however is too faint to provide
any measurable perturbation.
The star formation in NGC\,1156 is spread out in \hii\ regions
spanning the entire bar$+$disk, and the kinematics of the
ionized (and atomic) gas and stars is consistent with circular
rotation about a common axis \citep{hunter02}.

%\pagebreak
%\subsubsection*{NGC\,3353 (Haro\,3, Mrk\,35)}
\medskip
\noindent
{\subjectfont\selectfont $\star$~NGC\,3353 (Haro\,3, Mrk\,35)}

\smallskip
\noindent
NGC\,3353, the most metal-rich galaxy in our sample 
(\logoh$\sim$8.4, see Table \ref{tab:sample}),
is a dwarf irregular hosting a galaxy-wide starburst.
However, high-resolution radio images show that most of the starburst activity is 
concentrated within a single knot to the northwest of the galaxy nucleus
\citep{johnson04}.
The star clusters in this knot are quite young (1-4\,Myr) and massive
($\sim10^6$\,\msun) and reside
in a region with \av $\sim$2-8 mag, implying that much of the current 
star formation is currently enshrouded in dust \citep{johnson04}.
In fact, optical measures of the extinction are smaller than those
measured with the radio and near-infrared observations
\citep{steel96,cairos07}. 
Detailed photoionization models of the infrared fine-structure lines
suggest that there are at least two emission zones in this galaxy:
one with low extinction, visible in the optical, and another,
optically invisible with much higher extinction \citep{hunt06}.
Rotational \htwo\ has been detected in NGC\,3353, and
the detection of the high-excitation \oiv\ line ($\ga$4\,Ryd) implies
that radiative shocks are also present \citep{hunt10}.
Like other galaxies in this sample, kinematic evidence suggests that
a merger is responsible for triggering the starburst in NGC\,3353
\citep{cairos07}.

%\bigskip % to enforce attachment of title to paragraph
%\subsubsection*{NGC\,7077 (Mrk\,900)}
\medskip
\noindent
{\subjectfont\selectfont $\star$~NGC\,7077 (Mrk\,900)}

\smallskip
\noindent
NGC\,7077 is classified as an early-type galaxy, E/S0, but
with peculiar morphology, and relatively blue colors
\citep{gildepaz03,micheva13}.
It is an isolated galaxy, part of the \citet{vanzee00} sample,
with red colors in its outer regions (see also Fig. \ref{fig:images}).
These red colors are
in contrast with the strong (blue) nuclear starburst which has 
an intensity comparable to that of 30\,Doradus
\citep{vanzee00}.
Perhaps the most salient feature of NGC\,7077 is its extended
ultraviolet disk, suggesting that the disk of this galaxy is in
an active growth phase \citep{moffett12}.

%\subsubsection*{SBS\,0035-052}
\pagebreak
\medskip
\noindent
{\subjectfont\selectfont $\star$~\sbs}

\smallskip
\noindent
Like, \iizw,
we did not observe this source in \coone, but it is included
in our CO survey of higher-$J$ lines (Hunt et al., in prep.);
thus we consider it here.
This is the most metal-poor galaxy of our sample (\logoh$\sim$7.2),
with a neighbor at $\sim$22\,kpc distance 
\citep[SBS\,0335$-$052W,][]{pustilnik01}.
This companion, embedded in a common \hi\ cloud,
is the most metal-poor star-forming galaxy known in the
Local Universe \citep[\logoh$\sim$7.1,][]{lipovetsky99,izotov05}.
High-resolution optical and near-infrared imaging shows that the star 
formation activity in %\sbs(E) 
\sbs\ is confined to six massive SSCs, 
with the two youngest ones ($\la$3\,Myr) each comprising $\sim$5000 O7.5\,V stars
\citep{reines08}.
Despite its extremely low abundance, \sbs\ is known to 
have extremely high extinction \citep[\av$\ga$15\,mag,][]{plante02,houck04},
and $\ga 10^4$\,\msun\ of dust surrounding only the two youngest SSCs
\citep{hunt01,hunt14}.
Warm \htwo\ is clearly present in this galaxy, with
strong emission from ro-vibrational transitions at 2\,\micron\ 
\citep{vanzi00,vanzi11}.
The cm-regime radio spectrum is highly self-absorbed,
implying ionized-gas densities of $\ga$3000\,\cmthree\ \citep{hunt04,johnson09}.
\sbs\ is the proto-typical ``active'' starburst with its star-formation
activity occurring in compact (unresolved at {\it HST} resolution), dense
regions \citep{hirashita04}.

%\subsubsection*{UM\,448}
\medskip
\noindent
{\subjectfont\selectfont $\star$~UM\,448 (Arp\,161, UGC\,6665)}

\smallskip
\noindent
UM\,448 is a merger remnant with residual tidal tails, and so IR luminous ($L_{\rm IR}\,=\,6.3\times10^{10}$\,\lsun)
that it would almost be classified as a Luminous IR Galaxy \citep[LIRG,][]{dopita02}.
Despite its low metal abundance (\logoh$\sim$8.0, 0.2\,\zsun, see Table \ref{tab:sample}), this galaxy shows
clear PAH emission \citep{galliano08}.
This is perhaps because of a metallicity gradient which reaches a peak
value of \logoh$\sim$8.4 toward the northwest, a region of lower temperature \citep{james13}.
There is evidence for age gradients in the stellar populations and complex ionized emission-line kinematics,
both of which probably reflect the merger nature of UM\,448 \citep{james13}.

%\bigskip % to enforce attachment of title to paragraph
%\subsubsection*{UM\,462}
\medskip
\noindent
{\subjectfont\selectfont $\star$~UM\,462 (Mrk\,1307, UGC\,6850)}

\smallskip
\noindent
Like many of the galaxies in our sample, UM\,462 is possibly tidally interacting
with a relatively close neighbor, UM\,461
\citep{taylor95}, although \citet{vanzee98} concluded that 
the \hi\ kinematics are inconsistent with this.
Most of the star formation in UM\,462 occurs in two main knots of star formation, 
although high-resolution observations resolve at least four starbursting
regions with disrupted kinematics \citep{vanzi02,james10}. 
The Wolf-Rayet content of this galaxy has been disputed by \citet{james10}
who do not find evidence to support a significant Wolf-Rayet classification.
\citet{vanzi02} found ro-vibrational \htwo\ emission in this galaxy, but
no evidence for \feii, implying that the star-formation episodes must be
younger than $\sim$10\,Myr.
The knots of star formation are so compact and luminous as to be classified
as SSCs \citep{vanzi03}. 

\subsection{An additional source, I\,Zw\,18 \label{sec:izw18}}

Although \izw\ was not included in our sample, it was observed,
although not detected, in the \coone\ transition by \citet{leroy07}.
Because of its extreme low metallicity, 
similar to \sbs, we have incorporated this galaxy in
our analysis \citep[see also][]{schruba12}.
We have adopted the SFR from \citet{hunt05}, and calculated the stellar
mass, \mstar, as described below \citep[see also][]{fumagalli10}.

Like \iizw, \izw\ was discovered as an extragalactic \hii\ region by
\citet{sargent70}.
Together with the \sbs\ and SBS\,0035$-$052W pair, it is the most metal-poor
star-forming galaxy in the Local Universe with \logoh\,=\,7.17-7.22
\citep{thuan05}.
Most, if not all, of the current star formation occurs in two main clusters,
although the ``C component'' (or Zwicky's flare) is also forming
stars \citep{izotov01}.
However, the two clusters are not as extreme as those in \sbs\ (and other
galaxies in our sample), being somewhat 
less luminous than 30\,Doradus \citep{hunter95}.
Like \sbs, around the main star clusters
there is a complex structure of filaments and arcs from ionized
gas, and the complex \hi\ morphology and kinematics suggest that the
current star-formation episode was triggered by an interaction with
the C component \citep{lelli12}.
Unlike \sbs, however, there is no evidence for a dense gas component
in \izw; the radio continuum spectrum follows the 
trend expected for optically thin free-free emission \citep{hunt05} with
an ionized-gas density of $\la$100\,\cmthree, similar to
that found from optical spectra \citep{izotov01}.
Despite their similarly low metal abundance, the dust masses, dust-to-gas,
and dust-to-stars mass ratios in \izw\ are more than 100 times lower than in \sbs\
\citep{hunt14}, implying that the density of the cool gas in the ISM
may play an important role in dust formation \citep{schneider15}.

\subsection{SFRs and stellar masses}
\label{sec:sfrs}

We have calculated the SFRs for our targets according to \citet{calzetti10}
using \ha\ and 24\,\micron\ luminosities (their Eqn. 17).
This approach obviates the need for uncertain extinction corrections, and has
the added advantage of accounting for the general behavior of dust opacity in galaxies.
Many of the \ha\ total fluxes come from \citet[][Haro\,3, \iizw, Mrk\,996, NGC\,7077]{gildepaz03},
but others come from 
\citet[][\cgcg]{vanzee00},
\citet[][NGC\,1140, NGC\,1156]{kennicutt09},
\citet[][UM\,448]{dopita02}, and
\citet[][UM\,462]{james10}.
%When no total \ha\ measurements were available, for the
%most compact object we adopted \ha\ measurements from
%optical spectra \citep[][\cgcg, UM\,462:]{dopita02,kniazev04}.
The 24\,\micron\ fluxes are taken from \citet{engelbracht08},
\citet[][NGC\,7077]{temi09}, or
Hunt et al. (2015, in prep.).
For \sbs\ and \izw, we take the SFR measured by high-frequency radio free-free emission
\citep{johnson09,hunt05}.

Following \citet{grossi15},
stellar masses, \mstar, were estimated according to the recipe given by \citet{wen13}
based on WISE W1 (3.4\,\micron) luminosities.
This approach exploits the relatively constant mass-to-light ratios
of stellar populations, independently of metallicity and age, at near-infrared wavelengths 
\citep{norris14,mcgaugh14}.
However, when W1 photometry was not available, we used
IRAC 3.6\,\micron\ photometry instead.
In fact, the two bands are very similar;
using data from \citet{brown14}, \citet{grossi15} find for spirals a 
mean flux ratio $F_{3.4}/F_{3.6}\,=\,1.02\pm0.035$.
Including also the data for dwarf irregulars from \citet{brown14} gives a 
mean flux ratio $F_{3.4}/F_{3.6}\,=\,0.98\pm0.061$.
Thus, we conclude that the ratio of the W1 and IRAC 3.6\,\micron\ bands is
unity, with 5-6\% scatter for galaxies like our targets.

We used the \hii-galaxy formulation by \citet{wen13}, rather than what
they found for their full sample; the \hii\ galaxies have the lowest mass-to-light
ratios in their compilation, corresponding roughly to the bluest regions of
the galaxies studied by \citet{zibetti09}. 
To better take into account the weak trends with abundance found by \citet{wen13}, 
we also applied an approximate correction 
for low metallicity \citep[by multiplying the mass-to-light ratio
by 0.8:][see their Fig. 17]{wen13}.

However, before applying the \hii-galaxy relation by \citet{wen13}, we first
subtracted nebular emission and emission from hot dust.
Such contamination can be very important in the near-infrared
and can contribute 50\% or more to the observed flux at these wavelengths
\citep{hunt01,hunt02,smith09,hunt12}.
The ionized gas continuum contribution to the 3.4-3.6\,\micron\ flux was 
estimated from the SFR using the emission coefficients from \citet{osterbrock06}.
The subtraction of hot dust is based on the assumption that $H$-band emission
is entirely stellar.
Because $H$-band photometry is available for most of our sample, we used the data
from \citet{brown14} to estimate the maximum possible IRAC 3.6\,\micron/$H$-band
ratio in galaxies similar to our targets; 
95\% of the spiral/dwarf irregular galaxies have a flux ratio $\leq$2.4.
This corresponds to a (Vega-based) [H-3.6] color of $\sim0.8$, consistent
with what is found for the pure stellar component in star-forming galaxies
\citep{hunt02}.
After subtraction of the nebular component, any excess over this ratio was
attributed to hot dust and subtracted.
We found that hot dust can contribute between 10\% and 50\% in the 3.4-3.6\,\micron\
bands (most significantly for \iizw, \sbs). % and up to 80\%--90\% at 4.5\,\micron\ (for \iizw, \sbs).

We compared the stellar masses obtained with the formulation of 
\citet{wen13} to those calculated according to 
\citet{lee06} based on IRAC 4.5\,\micron\ luminosities \citep[used by][]{hunt12}, and find they are consistent
to within 0.2\,dex. 
Moreover, for the galaxies having both W1 and IRAC data, the stellar masses
obtained from 3.6\,\micron\ luminosities are within $\sim$5\% of those
from W1 as expected.
The Kroupa-like broken power-law initial mass function (IMF) 
used by \citet{calzetti10} and consequently for our calculation of SFRs is consistent with 
the IMF for \mstar\ by \citet{wen13}. 

\subsection{\hi\ measurements \label{sec:hi}}

All the sample galaxies have been observed in \hi, half of them with interferometric maps.
For the galaxies with resolved observations, for comparison with CO, we have
attempted to compute the \hi\ mass, \mhi, by integrating the mean \hi\
surface density within only the optical radius.
Because virtually all metal-poor dwarf galaxies show \hi\ emission well outside
their optical radius
\citep[e.g.,][]{hunter94a,vanzee98,vanzee00,ekta09,lelli12,lelli14},
these values of \mhi\ are typically a factor of 2-10 lower than the total \hi\ mass,
integrated over the entire \hi\ extent.
Table \ref{tab:sample} reports both values of \mhi\ when resolved \hi\ maps
are available.

\subsection{Comparison samples \label{sec:comparison}}

To expand the parameter space for our analysis, we also considered
another dwarf sample with CO detections \citep{schruba12},
as well as normal star-forming galaxies, LIRGs, and
Ultra-LIRGs \citep[ULIRGs,][]{gao04,graciacarpio08,garciaburillo12}.
For the latter galaxies, we converted their published IR luminosity to SFR
according to \citet{kennicutt98}, after applying the correction
to a \citet{kroupa01} IMF for consistency
with our SFRs and those of \citet{schruba12}. 
For the Schruba et al. CO(2--1) observations,
we applied the factor they recommend to convert \cotwo\ to \coone:
$I_{\rm CO2-1}/I_{\rm CO1-0}\,=\,0.7$.
Stellar masses for the \citet{gao04} and \citet{garciaburillo12} samples
were taken from \citet[][]{skibba11},
and for the galaxies from \citet{schruba12}, masses were adopted either from \citet{hunt12} or \citet{lee06}.
These latter estimates are based on IRAC photometry similar to the method used here, as
described in Sect. \ref{sec:sfrs}.

We incorporated also the galaxies 
from the CO(2--1) APEX Low-redshift Legacy Survey of MOlecular Gas
\citep[ALLSMOG,][]{bothwell14}.
This sample is essentially an extension to lower stellar masses (8.5\,$<$\,log(\mstar/\msun)\,$\leq$\,10)
of the COLDGASS galaxies studied by \citet{saintonge11a,saintonge11b}.
For consistency with \citet{schruba12}, to convert \cotwo\ to \coone, we 
again use $I_{\rm CO2-1}/I_{\rm CO1-0}\,=\,0.7$, rather than
$I_{\rm CO2-1}/I_{\rm CO1-0}\,=\,1$ as in \citet{bothwell14}.
Because of the $\sim$0.4\,dex offset between the ALLSMOG metallicity calibration
and the mainly direct-temperature\footnote{The electron temperature
derived from the ratio of the \oiii\,$\lambda$4363 auroral line to lower-excitation lines
(\oiii\,$\lambda$4959,\,5007) is converted to a metallicity after correcting
for unseen phases of ionization \citep[e.g.,][]{osterbrock06}.}
determinations used here \citep[e.g.,][]{hunt12,andrews13},
the oxygen metallicities given by \citet{bothwell14} were converted
to the \citet{pettini04} calibration according to \citet{kewley08}\footnote{These metallicites
are shown in Figs. \ref{fig:sfrco_oh}, \ref{fig:ssfroh}, \ref{fig:cosfr_res}, 
\ref{fig:tdep}, \ref{fig:h2frac}, \ref{fig:gasfrac}, and \ref{fig:tdephi}.}.
Finally, to extend the metallicity range to extremely metal-poor galaxies (XMPs),
we incorporate the sample from \citet{filho13} which, however, have no CO measurements;
we include them to better investigate gas content at low metallicities.

Both the ALLSMOG and the \citet{filho13} galaxies have stellar masses and SFRs
taken from the MPA-JHU catalogue \citep{kauffmann03,brinchmann04},
the same basis for calibration used by \citet{wen13}.
However, the automatic algorithm used by the SDSS does not
always provide the correct global photometry for galaxies whose surface brightness
is not uniformly distributed (e.g., in clumps or knots).
This should not be a problem for the ALLSMOG galaxies, which tend to be
extended spirals, but could be a problem for the comparison with the XMP sample
by \citet{filho13}.
%\citet{filho13} also find that the discrepancy tends to be 
In fact, \citet{filho13} found that comparing mass-to-light ratios based on
optical colors gave as much as an order of magnitude \mstar\ excess over the SDSS estimates;
the discrepancies tend to be 
largest for galaxies with either cometary or two-/multiple-knot morphology.
To check for systematic differences between the near-infrared
method used here (IRAC, WISE) and in \citet{hunt12} \citep[see also][]{madden13}
and the optically-determined (MPA-JHU) SDSS values, we searched the MPA-JHU
database and found 17 low-metallicity galaxies for which we had calculated
stellar masses in \citet{hunt12} using 4.5\,\micron\ luminosities.
Although the contributions of the nebular continuum and hot dust emission have
been subtracted,
the values of \mstar\ using IRAC/WISE photometry are, on average, 17 times larger than
those with SDSS with a scatter of $\sim$0.4\,dex.
We investigated the morphology of the galaxies in this comparison and find that
they tend to be very clumpy, with either multiple knots or cometary features.
Comparing the SDSS photometry with the broad-band $JHK$ photometry for these galaxies gives 
NIR-optical colors that are extremely red, unrealistically so for these blue galaxies; the 
implication is that the optical fluxes are being underestimated.
We thus ascribe this difference to the difficulty of the SDSS automated
photometric algorithms when confronted with multiple clumps.
As a consequence, in our comparisons with the XMP sample, 
these uncertainties should be kept in mind.

\section{The CO observations \label{sec:observations}}

We have observed \coone\ %, \cotwo, \cothree, 
in our sample galaxies over a three-year period from 2008 to 2010
with the IRAM 30-m telescope (Pico Veleta, Spain). 
%The measurements of \cothree\ were acquired with the APEX 12-m telescope (Chile),
%and the remaining transitions with the IRAM 30-m telescope (Pico Veleta, Spain).
%The higher-J transitions are discussed in a companion paper; here we focus
%on \coone.

\subsection{IRAM observations \label{sec:iram}}

In September, 2009, and September, 2010, we 
observed six galaxies\footnote{\iizw\ and \sbs\ were observed only in the higher-J lines
with APEX.} in \coone\ with the Eight Mixer Receiver (EMIR)
using the Wideband Line Multiple Autocorrelator (WILMA) backend. 
Observations of \coone\ for NGC\,1140 and NGC\,1156
were performed with the AB receivers in November, 2008, using the 1 and
4\,MHz backends at 3 and 1.3\,mm (88-115\,GHz, 219-229\,GHz, respectively).
The precipitable water vapor was $\la$4\,mm during the 2008 run, 
giving system temperatures of T$_{\rm sys}$ of $\la$370\,K %(270--330\,K) 
at 3\,mm. %(1.3\,mm). 
In 2009, the conditions were worse ($\la$11\,mm), but
the EMIR receivers had comparable T$_{\rm sys}$, $\sim$335\,K %(330\,K) 
at 3\,mm; %(1.3\,mm)
in 2010, our most extensive run, T$_{\rm sys}$ varied from $\sim$230\,K
to $\ga$400\,K, with precipitable water vapor ranging from $\sim$ 4\,mm to 10\,mm.
%For NGC\,1140,
%we used the EMIR E090 receiver tuned to an intermediate frequency, $\sim$113\,GHz,
%in order to cover other 3\,mm transitions simultaneously together with \coone\
%in the lower-inner sideband in two polarizations.
In all runs, we used wobbler switching with a throw of 90\arcsec.
Intensity calibration was performed with two absorbers at different temperatures. 
 
Pointing was checked at regular intervals (every 1.5\,hr) on nearby planets %(Uranus)
or bright quasars (e.g., 0234$+$285), and focus tests
were carried out every 4\,hrs during the night and every 3\,hrs 
during the day.
Pointing errors never exceeded $\sim$3\arcsec, and were usually $\la$2\arcsec.
Given the beam size (half-power beam width) of $\sim$22\,\arcsec,
this was more than sufficient.

\begin{figure*}[ht]
\begin{minipage}[t]{\textwidth}
\hbox{
\centerline{
\includegraphics[angle=0,width=0.3\linewidth]{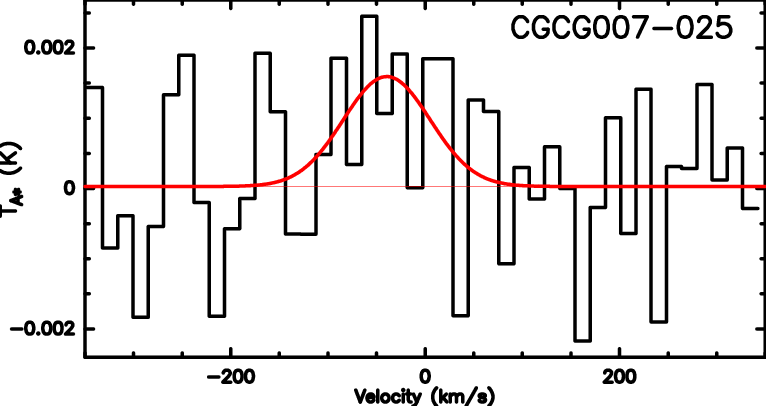}
\hspace{0.2cm}
\includegraphics[angle=0,width=0.3\linewidth]{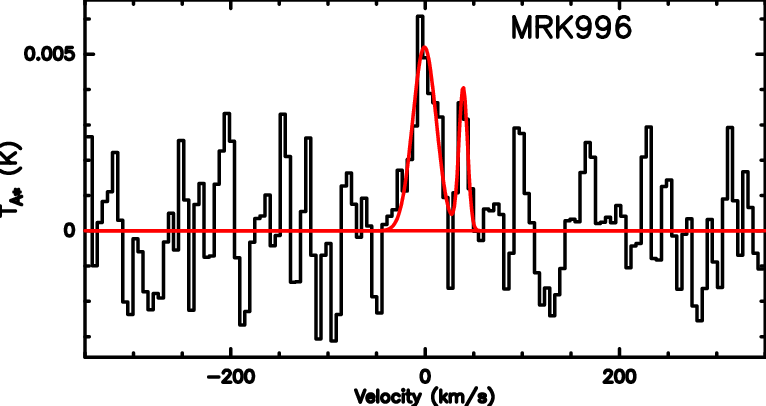}
\hspace{0.2cm}
\includegraphics[angle=0,width=0.3\linewidth]{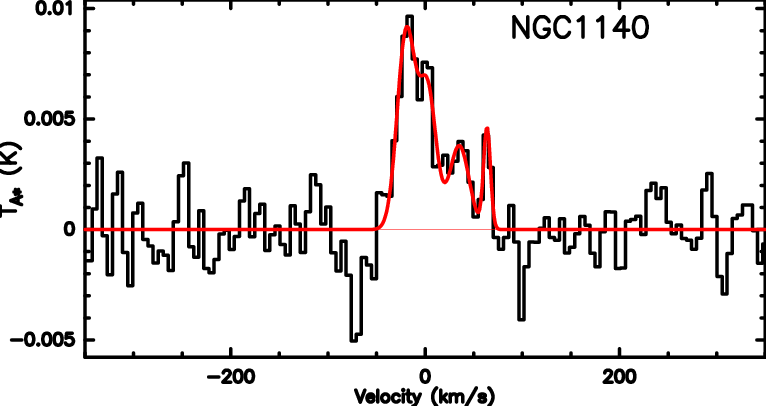}
}
}
\vspace{\baselineskip}
\hbox{
\centerline{
\includegraphics[angle=0,width=0.3\linewidth]{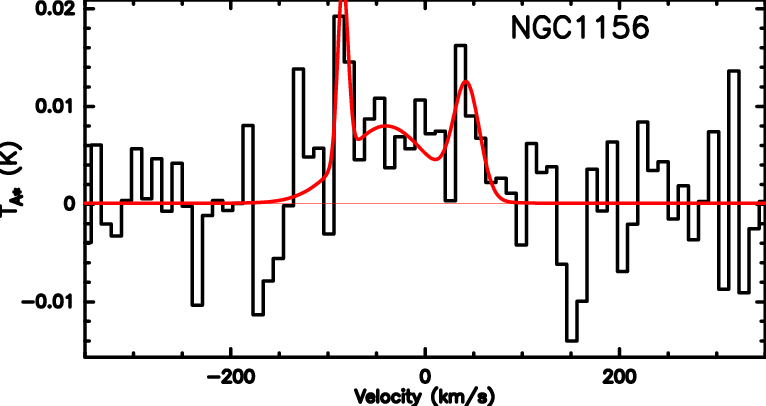}
\hspace{0.2cm}
\includegraphics[angle=0,width=0.3\linewidth]{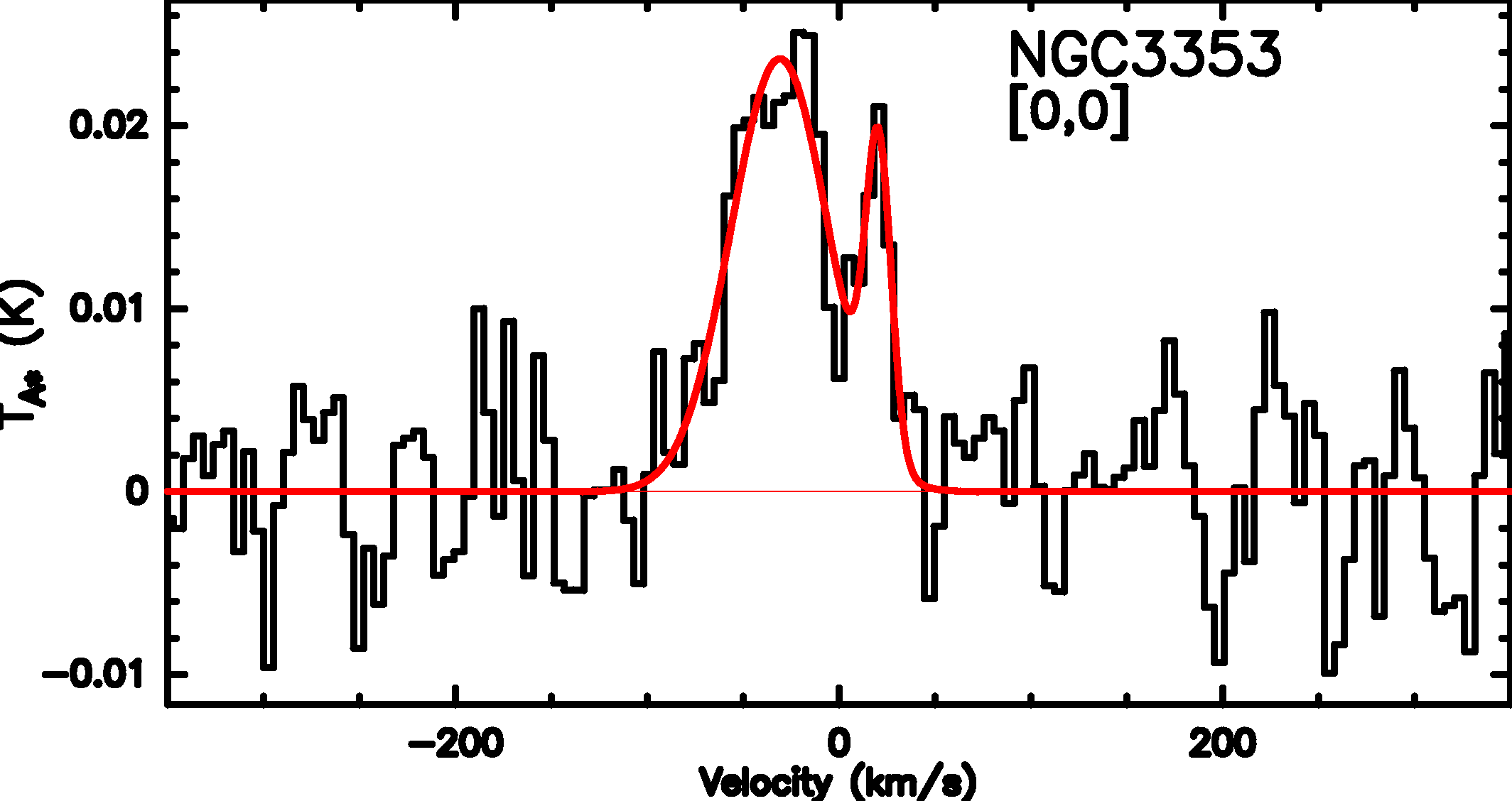}
\hspace{0.2cm}
\includegraphics[angle=0,width=0.3\linewidth]{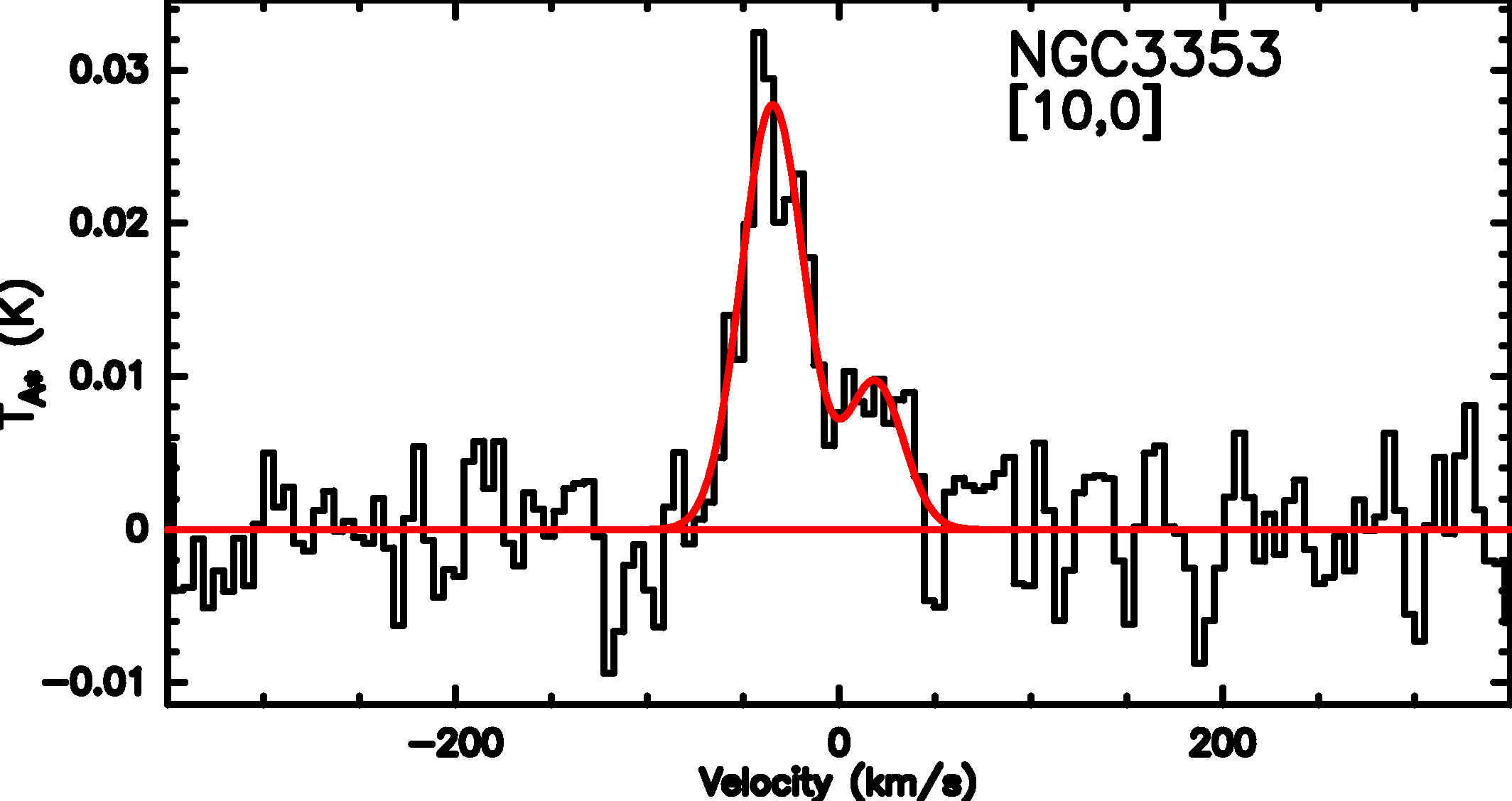}
}
}
\vspace{\baselineskip}
\hbox{
\centerline{
\includegraphics[angle=0,width=0.3\linewidth]{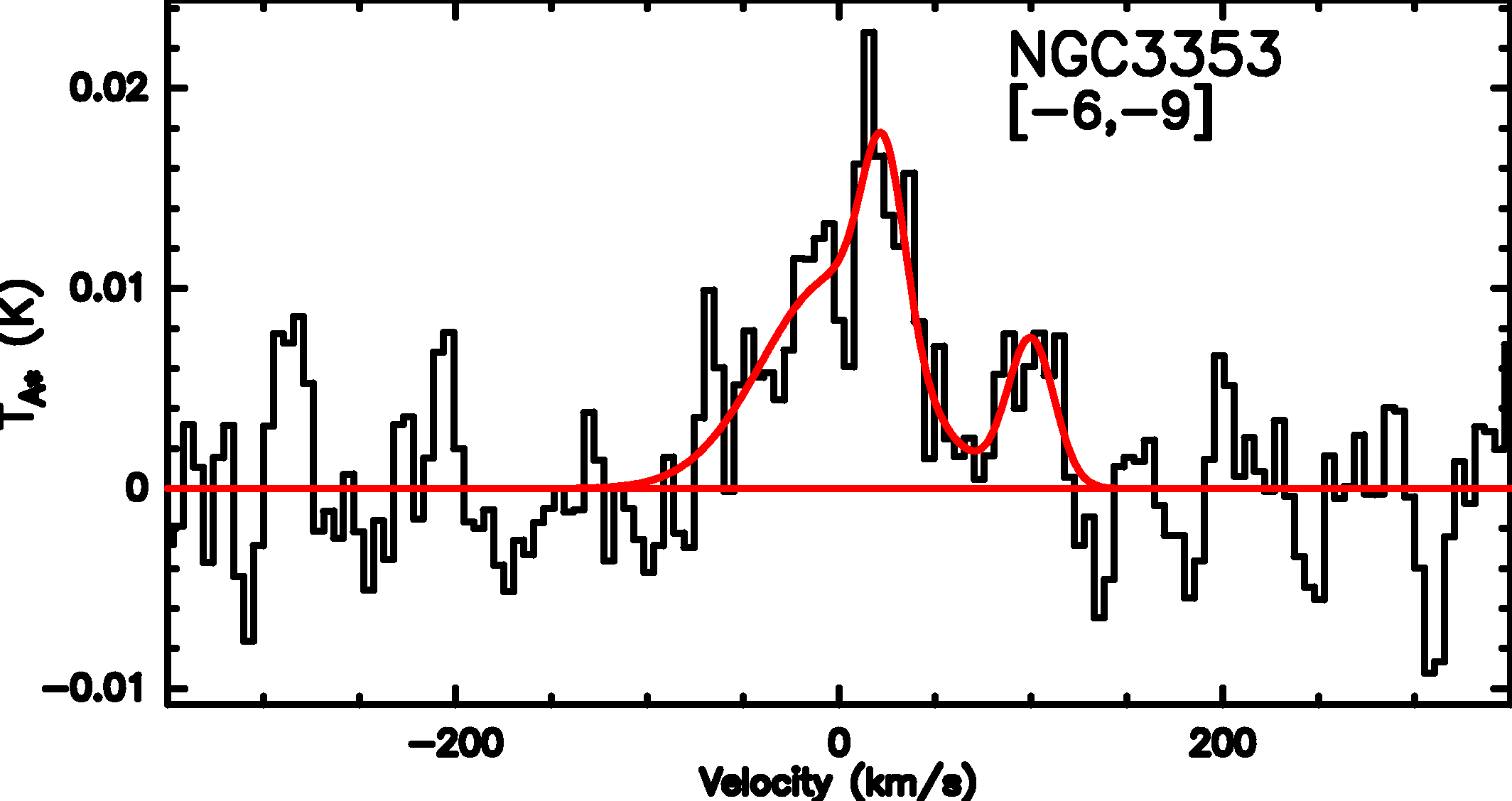}
\hspace{0.2cm}
\includegraphics[angle=0,width=0.3\linewidth]{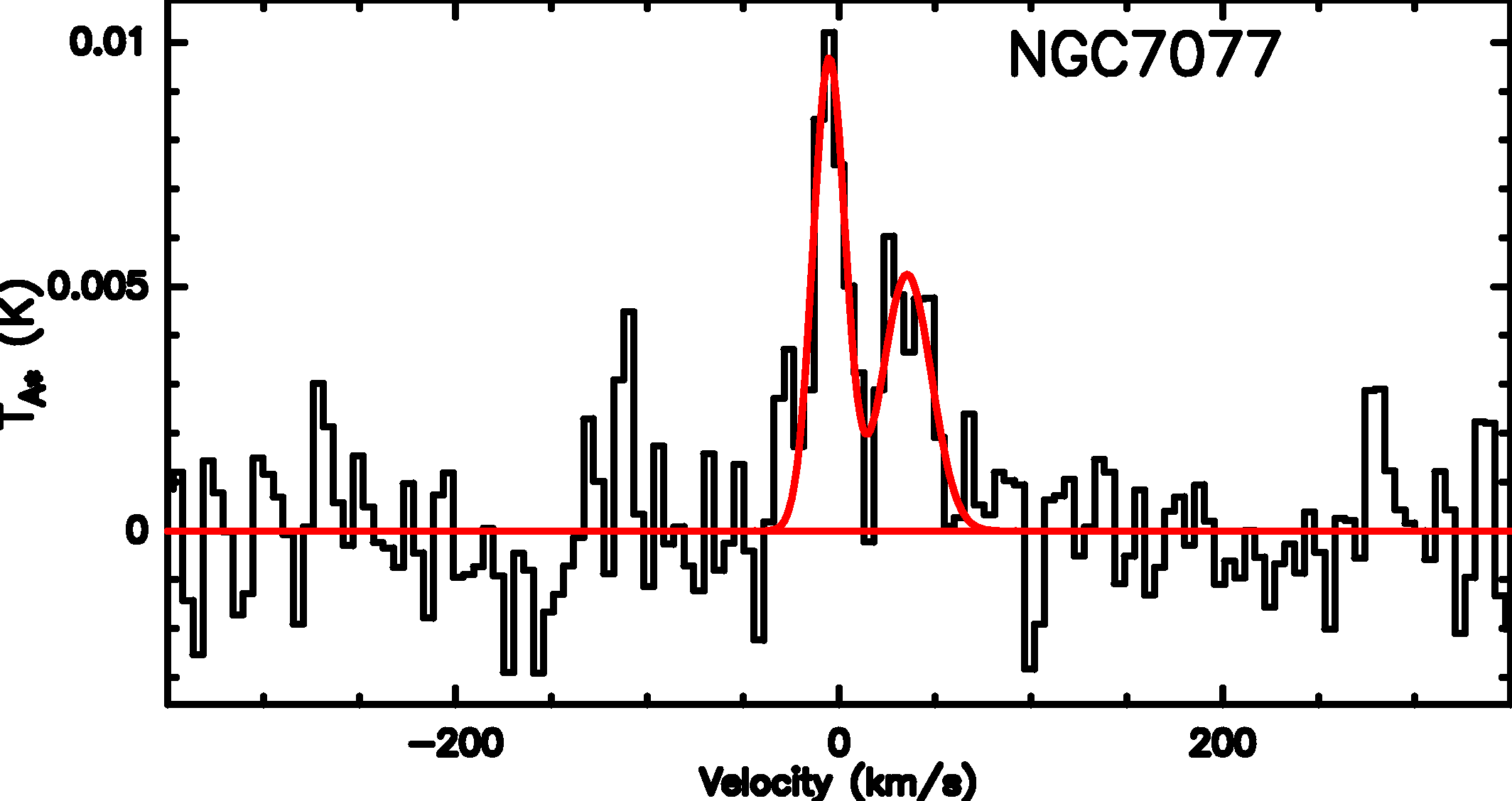}
\hspace{0.2cm}
\includegraphics[angle=0,width=0.3\linewidth]{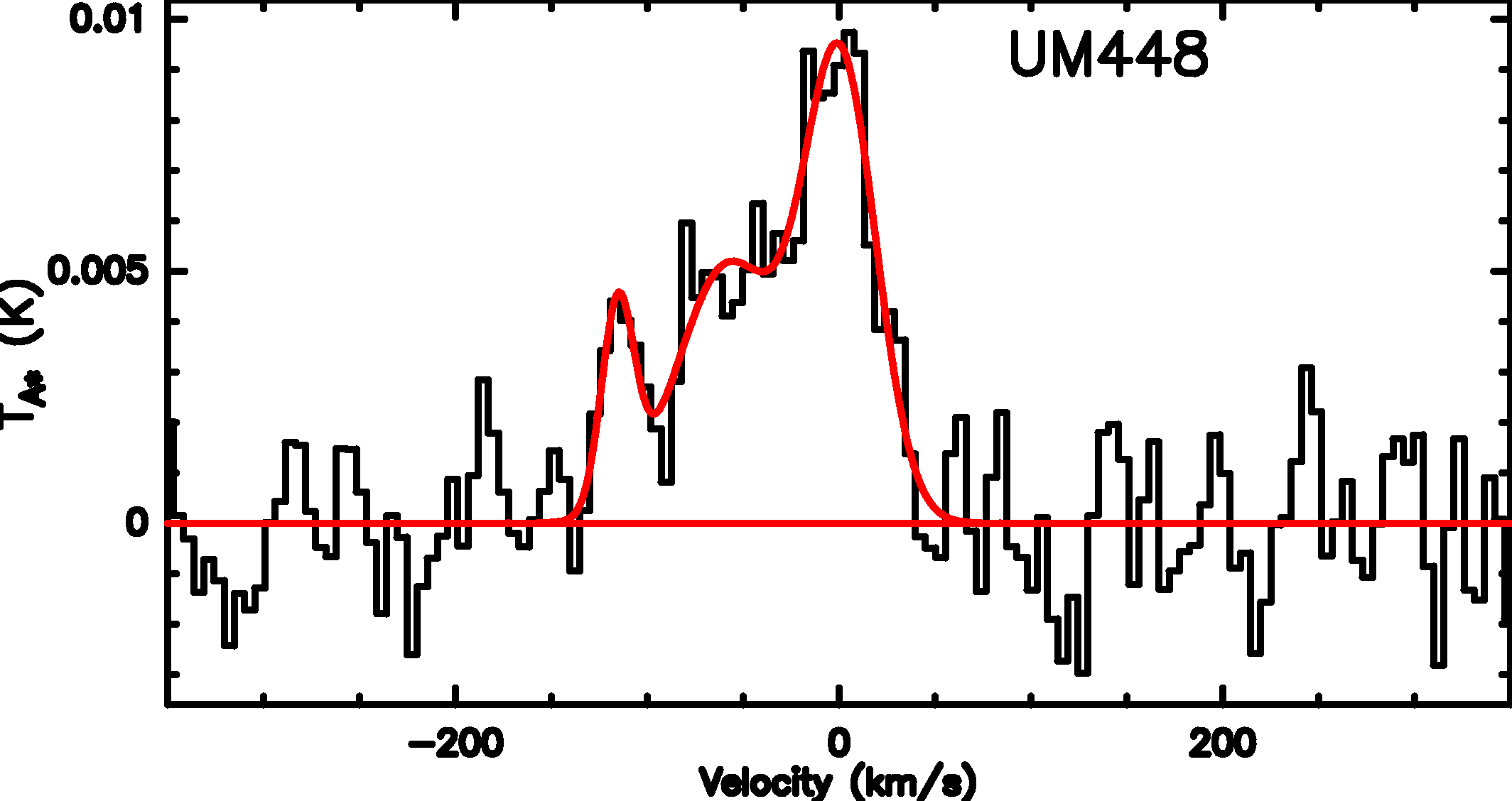}
}
}
\end{minipage}
%\end{figure*}
%\nopagebreak
%\setcounter{figure}{2}
%\begin{figure*}
\hbox{
\centerline{
\begin{minipage}[c]{0.3\textwidth}
\vspace{\baselineskip}
\includegraphics[angle=0,width=\linewidth]{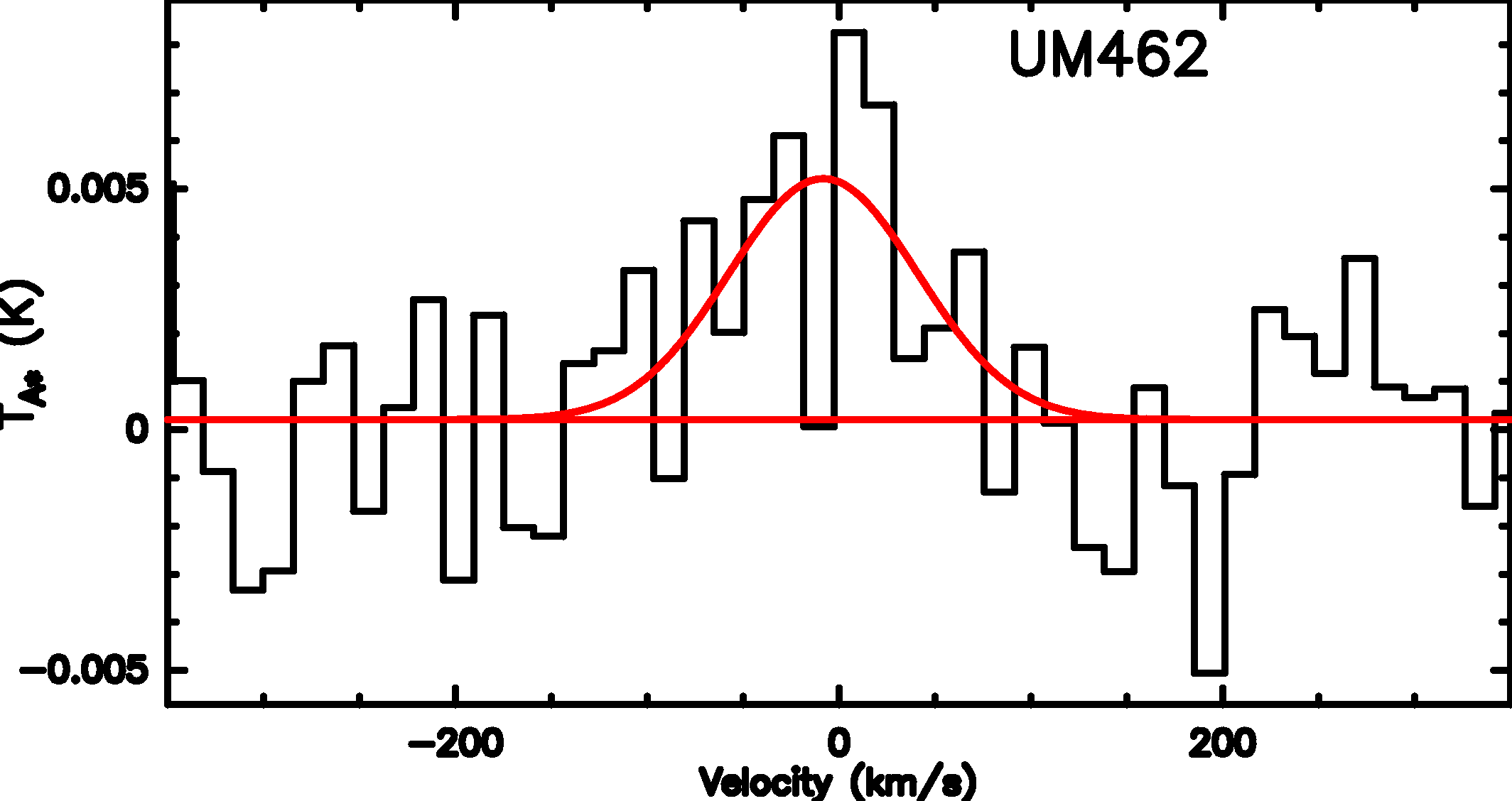}
\end{minipage}
\hspace{0.8cm}
\begin{minipage}[c]{0.585\textwidth}
\vspace{-0.8cm}
\caption{\coone\ average spectra for the observed galaxies.
Top row, from left to right in alphabetical order of name: 
\cgcg,  Mrk\,996, and NGC\,1140.
Second row: NGC\,1156, NGC\,3353 (Haro\,3) two pointings
([0,0], [0,+10\arcsec].
Third row: 
NGC\,3353 third pointing  [-6\arcsec, -9\arcsec]),
NGC\,7077 (Mrk\,900), UM\,448.
Bottom row: UM\,462.
The baselines are shown as a horizontal solid line, together with the multiple-component
Gaussian fits as described in Sect. \ref{sec:reduction}.
The vertical axes are in \ta\ units, while in Table \ref{tab:lines} the units are \tmb,
converted from \ta\ as described in the text.
Velocity channels are as reported in Table \ref{tab:lines}.
\label{fig:lines}
}
\end{minipage}
}
}
\end{figure*}

% -----------------------------------------------------------------
% --- Table 2: observations
%
\begin{center}
\begin{table*}
      \caption[]{\coone\ emission line parameters} 
\label{tab:lines}
%\resizebox{\linewidth}{!}{
%\addtolength{\tabcolsep}{7pt}
{\small
%\tiny
\begin{tabular}{lrcrrrrr}
\hline
\multicolumn{1}{c}{Name} &
\multicolumn{1}{c}{Telescope time} &
\multicolumn{1}{c}{Component} &
\multicolumn{1}{c}{Channel} &
\multicolumn{1}{c}{T$_{\rm rms}$(mb)$^{\rm a}$} &
\multicolumn{1}{c}{Offset} &
\multicolumn{1}{c}{FWHM} &
\multicolumn{1}{c}{Integrated intensity$^{\rm b}$} \\
& \multicolumn{1}{c}{(min)} & &
\multicolumn{1}{c}{(\kms)} &
\multicolumn{1}{c}{(K)} &
\multicolumn{1}{c}{(\kms)} &
\multicolumn{1}{c}{(\kms)} &
\multicolumn{1}{c}{(\kkms)} \\
\hline
\\
CGCG007-025                          & 524 & Total                   & 10.5 & 0.0018 & -39 & 115 & 0.286 (0.095) \\
                                     &     & Total                   & 15.7 & 0.0016 & -39 & 105 & 0.285 (0.102) \\
\\
\iizw$^{\rm c}$                      & \multicolumn{1}{r}{$-$}       \\
\\
Mrk\,996                             & 573 & 1                       &  5.2 & 0.0019 &  -1 &  28 & 0.18 (0.03) \\ 
                                     &     & 2                       &  5.2 &        &  39 &  10 & 0.05 (0.02) \\ 
                                     &     & Total                   &  5.2 &        &   0 &  \multicolumn{1}{r}{$-$} & 0.23 (0.04) \\ 
\\
%NGC\,1140                            & 831$^{\rm d}$ & 1             &  5.2 & 0.0016 & -17 &  26 & 0.32 (0.05) \\ 
NGC\,1140                            & 831 & 1                       &  5.2 & 0.0016 & -17 &  26 & 0.32 (0.05) \\ 
                                     &     & 2                       &  5.2 &        &   3 &   8 & 0.07 (0.04) \\ 
                                     &     & 3                       &  5.2 &        &  30 &  38 & 0.18 (0.05) \\ 
                                     &     & 4                       &  5.2 &        &  64 &   7 & 0.04 (0.02) \\ 
                                     &     & Total                   &  5.2 &        &     &  \multicolumn{1}{r}{$-$} & 0.61 (0.08) \\ 
                                     &     & Total$^{\rm d}$         &  5.2 &        &   0 &  \multicolumn{1}{r}{$-$} & 1.00 (0.13) \\ 
\\
NGC\,1156                            &  47 & 1                       & 10.4 & 0.0076 & -84 &  13 & 0.36 (0.09) \\
                                     &     & 2                       & 10.4 &        & -35 & 100 & 0.89 (0.09) \\
                                     &     & 3                       & 10.4 &        &  44 &  42 & 0.48 (0.09) \\
                                     &     & Total                   & 10.4 &        &   0 & \multicolumn{1}{r}{$-$} & 1.25 (0.16) \\
                                     &     & Total$^{\rm d}$         & 10.4 & 0.0076 &   0 & \multicolumn{1}{r}{$-$} & 4.04 (0.52) \\
\\
NGC\,3353 [0\arcsec, 0\arcsec]       &  79 & 1                       & 5.2  & 0.0056 & -31 &  60 & 1.83 (0.17) \\
                                     &     & 2                       & 5.2  &        &  21 &  16 & 0.34 (0.11) \\
                                     &     & Total                   & 5.2  &        &   0 &  \multicolumn{1}{r}{$-$} & 2.17 (0.20) \\
                                     &     & Total$^{\rm d}$         & 5.2  &        &   0 &  \multicolumn{1}{r}{$-$} & 3.07 (0.29) \\
NGC\,3353 [0\arcsec, $+$10\arcsec]   &  77 & 1                       & 5.2  & 0.0045 & -35 &  42 & 1.38 (0.12) \\
                                     &     & 2                       & 5.2  &        &  19 &  22 & 0.43 (0.12) \\
                                     &     & Total                   & 5.2  &        &   0 &  \multicolumn{1}{r}{$-$} & 1.81 (0.17) \\
NGC\,3353 [$-$6\arcsec, $-$9\arcsec] &  80 & 1                       & 5.2  & 0.0046 & -17 &  71 & 0.84 (0.20) \\
                                     &     & 2                       & 5.2  &        &  25 &  36 & 0.62 (0.18) \\
                                     &     & 3                       & 5.2  &        &  98 &  31 & 0.30 (0.08) \\
                                     &     & Total                   & 5.2  &        &   0 &  \multicolumn{1}{r}{$-$} & 1.46 (0.27) \\
\\
NGC\,7077                            & 585 & 1                       & 5.2  & 0.0017 &  -5 &  20 & 0.25 (0.03) \\
                                     &     & 2                       & 5.2  &        &  35 &  30 & 0.20 (0.04) \\
                                     &     & Total                   & 5.2  &        &   0 &  \multicolumn{1}{r}{$-$} & 0.45 (0.05) \\
\\
%\sbs$^{\rm e}$                       &  \multicolumn{1}{r}{$-$}  \\
\sbs$^{\rm e}$                       &  \multicolumn{1}{r}{$-$}  \\
\\
UM\,448                              & 454 & 1                       & 5.3  & 0.0017 & -115 &  20 & 0.11 (0.04) \\
                                     &     & 2                       & 5.3  &        &  -58 &  63 & 0.42 (0.16) \\
                                     &     & 3                       & 5.3  &        &    1 &  44 & 0.52 (0.14) \\
                                     &     & Total                   & 5.3  &        &    0 &  \multicolumn{1}{r}{$-$} & 1.04 (0.22) \\
\\
UM\,462                              & 119 & Total                   & 10.4 & 0.0034 &  -9 & 121 & 0.824 (0.173) \\
                                     &     & Total                   & 15.7 & 0.0032 &  -8 & 116 & 0.749 (0.207) \\
\\
\hline
\end{tabular}
%}
}
\vspace{0.5\baselineskip}
\begin{description}
\item
[$^{\mathrm{a}}$] This and all temperature units in this table are
main-beam, \tmb.
\item
[$^{\mathrm{b}}$] Values in parentheses are the uncertainties.
The total line flux is the sum of the individual components, and the
total uncertainty is calculated by adding in quadrature
the individual uncertainties. 
%The total line width is the quadrature sum of the individual widths. 
\item
[$^{\mathrm{c}}$] As discussed in the text, we did not observe
this galaxy in \coone, but rather relied on the data reported by
\citet{taylor98}.
\item
%[$^{\mathrm{d}}$] This long integration time was dictated by
%the requirement for \hcn, discussed in the companion paper.
[$^{\mathrm{d}}$] With aperture corrections as described in the text.
\item
[$^{\mathrm{e}}$] We did not observe
this galaxy in \coone, but rather adopted the \cothree\ data reported by
\citet{hunt14}, converted to CO(1--0) assuming a CO(3--2)/CO(1--0) flux ratio of 0.6. 
\end{description}
\end{table*}
\end{center}
% -----------------------------------------------------------------

\subsection{Data reduction \label{sec:reduction}}

We adopted the GILDAS/CLASS data reduction package to obtain mean spectra
(http://www.iram.fr/IRAMFR/GILDAS).
To remove the continuum from each spectrum, a polynomial baseline 
%of order 3-4 %order of 0 or 1 
was fitted to the line-free regions of each scan (defined a priori) and 
subtracted; the scans were thereafter smoothed and averaged, and a constant baseline subtracted. 
We measured the peak
intensities, central velocities, full width half-maximum (FWHM) and
velocity-integrated fluxes of the detected lines by fitting Gaussian profiles
to the data. 
In most cases, there is significant velocity structure in the emission, so
we fit the line profiles to multiple Gaussians.
In these cases we use the sum of the integrated line intensities 
from the multiple Gaussian fits in the analysis (see Table \ref{tab:lines}).

Antenna temperatures (\ta) have been converted to main-beam temperatures (\tmb)
by dividing the antenna temperatures by $\eta\,\equiv\,B_{\rm eff}/F_{\rm eff}$, 
where $B_{\rm eff}$ (=\,0.78) and $F_{\rm eff}$ (=\,0.94) are
the beam and forward hemisphere efficiencies, respectively 
(i.e., \tmb\,=\,\ta/$\eta$).  
To convert the measured \coone\ brightness temperatures,
[\tmb\ (K)] to fluxes [S (Jy)], we used the standard IRAM 30-m conversion factor
of 3.906\,Jy\,K$^{-1}$.
The \coone\ line profiles are shown in Fig. \ref{fig:lines}, together with the
Gaussian fits obtained using the GILDAS/CLASS package as
described above and reported in Table \ref{tab:lines}.
Although a few of the individual velocity components are too narrow to be significant
(e.g., NGC\,1140, NGC\,1156), we have checked that the total intensities 
are robust to the details of the fitting. 

\subsection{Aperture corrections \label{sec:aperture}}

The IRAM 30 m has a beam with an FWHM of $\sim$22\,\arcsec\ at 3\,mm.
Consequently, the largest galaxies in our sample
(NGC\,1140, NGC\,1156, NGC\,3353, see Table \ref{tab:sample})
may be missing flux.
We have attempted to correct for this by adopting the approach
of \citet{saintonge11a} who calibrated aperture corrections
using the set of nearby galaxies observed by \citet{kuno07}.
We applied Eqn. (2) by \citet{saintonge11a} based on
optical size, but used the geometric means of the major
and minor axes given in Table \ref{tab:sample}.
These corrections amount to an increase in line intensities
by a factor of 1.64 for NGC\,1140 and 3.23 for NGC\,1156.

For NGC\,3353 (Haro\,3), we have additional pointings spaced
by roughly 0.5 beams, hence not independent.
Where necessary, \citet{saintonge11a} used spacings of 0.75~beams
along the major axis, and found a mean ratio between the 
offset flux and the central flux of 0.33.
The corresponding ratio for the [-6\arcsec, -9\arcsec] pointing of NGC\,3353
(along the southern major axis) is 0.53, larger than the 
offset corresponding to the wider spacing by \citet{saintonge11a},
but consistent with what could be expected for the closer
spacing of our offset.
Hence, for NGC\,3353, we adopted Eqn. (3) of \citet{saintonge11a}
with their mean ratio between offset and central intensities (0.33);
this correction gives a multiplicative factor of 1.41.
Table \ref{tab:lines} also reports the \coone\ intensities
corresponding to the aperture corrections for these galaxies
which we will adopt throughout the paper.

\subsection{Comparison with previous CO observations \label{sec:cocomparison}}

Because some of the galaxies in our sample have previous CO observations,
here we compare our results with those.
\citet{sage92} observed NGC\,3353, NGC\,7077, UM\,448, and UM\,462,
and securely detected only the former two.
The results for NGC\,3353, obtained with the IRAM 30-m, are in good agreement with our central pointing
(they find an integrated CO intensity of 1.98$\pm$0.28\,\kkms), although
we find $\sim$70\% of their CO emission for NGC\,7077 (they find 0.68$\pm$0.1\,\kkms).
%Our formal uncertainty is rather small, so the reason for the discrepancy is unclear.
%\citep[although see][]{taylor98}.
Our detections of UM\,448 and UM\,462 are consistent with their marginal detections
obtained with the NRAO 12-m.

UM\,462 was also observed by \citet{brinks97} and \citet{gondhalekar98}, neither of
whom detected the galaxy; their upper limits ($<$0.28\,\kkms\ and $<$0.5\,\kkms, respectively)
are somewhat discrepant with our detection (0.7-0.8\,\kkms), 
as well as with the marginal detection of \citet[][$\sim$0.7\,\kkms]{sage92}.

NGC\,1140 was observed previously by \citet{albrecht04} and
NGC\,1156 by \citet{leroy05}.
%With similar velocity binning, %and considering the larger beam ($\sim$55\arcsec) of \citet{leroy05},
The two previous observations are reasonably consistent with our results. 

\section{CO luminosity and SFR \label{sec:analysis}}

Our new IRAM observations have successfully detected \coone\ down to 
a metallicity of \logoh\,=\,7.74 (\cgcg, signal-to-noise ratio 3.0, see Table \ref{tab:lines}),
which is the lowest metallicity at which CO has been (tentatively) 
detected %outside the Local Group
\citep[see WLM: \logoh\,=\,7.8,][]{elmegreen13}.
To our knowledge, Table \ref{tab:lines} gives the largest number of \coone\ detections
ever reported for \logoh$\la$8 (0.2\,\zsun) outside the Local Group.
 
\begin{figure}[!h]
\vspace{2\baselineskip}
\includegraphics[angle=0,width=\linewidth]{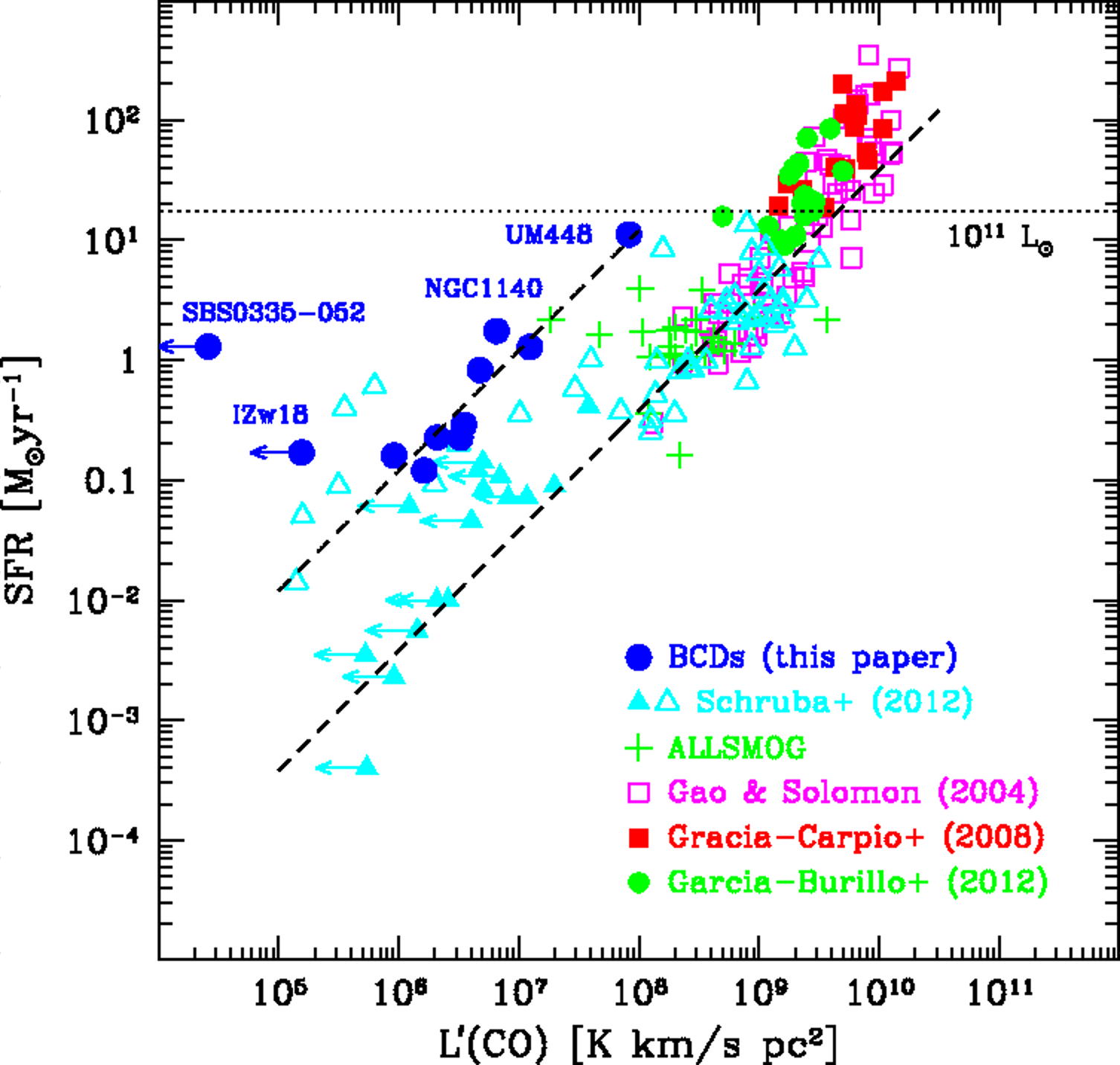}
\caption{SFR vs. \lco\ for our sample and additional galaxies from
the literature.
As discussed in Sect. \ref{sec:izw18},
we include \izw\ as one of our sample galaxies \citep[see also][]{schruba12}. 
The open triangles for \citet{schruba12} correspond to their compilation of literature data 
and the filled ones to their metal-poor dwarf observations.
CO non-detections are shown by leftward arrows.
The rightmost regression line corresponds to the linear correlation
found by \citet{gao04}, and the leftmost one to our new determination,
also compatible with a slope of unity (see text).
The horizontal dotted line gives the IR luminosity limit above which
the linear LIR-\lco\ correlation of \citet{gao04} no longer holds.
}
\label{fig:sfrco}
\end{figure}

In the following, we compare the integrated CO luminosity, \lco, with SFR as calculated
in Sect. \ref{sec:sfrs}.
\lco\ was derived according to the formulation of \citet{solomon05}, using
the appropriate Jy-to-K conversion (see Sect. \ref{sec:observations} for the IRAM value).
%For NGC\,3353 (Haro\,3), which we observed in multiple pointings, we use
%the sum of the three pointings. 
The result is shown in Fig. \ref{fig:sfrco}.

Not surprisingly, most of the metal-rich star-forming galaxies in the comparison
samples follow
the linear correlation between SFR and \lco, as formulated by \citet{gao04},
shown as the rightmost dashed line (corrected to the Kroupa IMF used here for SFR,
see Sect. \ref{sec:sample});
the horizontal dotted line in Fig. \ref{fig:sfrco}
gives the IR luminosity limit above which
the linear LIR-\lco\ correlation of \citet{gao04} no longer holds.
The metal-poor dwarf galaxies in our sample show a similar correlation but
offset to lower \lco\ by a factor of $\sim$30: 
$$\log({\rm SFR})\,=\,\log({\rm L}^\prime_{\rm CO}) - (6.92\pm0.07) .$$
The best-fitting slope obtained with a robust fitting technique
\citep{li85,fox97} is unity within the errors (1.05$\pm$0.15); thus,
for consistency with \citet{gao04}, we fixed a unit slope 
and derived the corresponding intercept as given above and shown in Fig. \ref{fig:sfrco}. 
Although we did not use the data of \citet{schruba12} to establish the best-fit
regression, most of those galaxies also follow the trend for our sample (see Fig. \ref{fig:sfrco}).
The exceptions are the two most metal-poor galaxies, \sbs\ and \izw.

\begin{figure}[!h]
\vspace{2\baselineskip}
\includegraphics[angle=0,width=\linewidth]{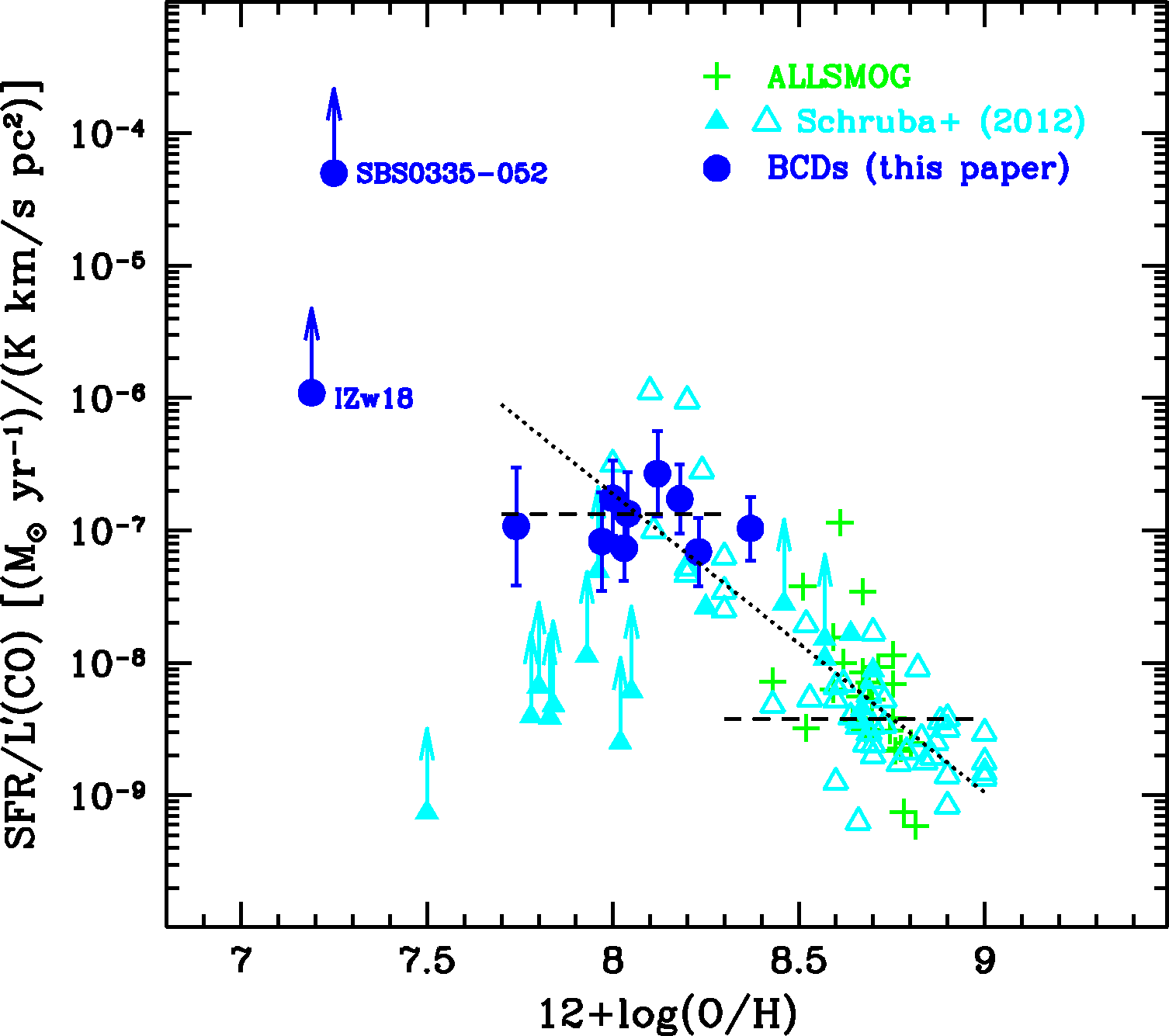}
\caption{SFR/\lco\ vs. \logoh\ for our sample and additional galaxies from
the literature.
As discussed in Sect. \ref{sec:izw18},
we include \izw\ as one of our sources \citep[see also][]{schruba12}. 
The open triangles for \citet{schruba12} correspond to their compilation of literature data 
and the filled ones to their metal-poor dwarf observations.
CO non-detections are shown by upward arrows.
The two horizontal dashed lines correspond to the regressions with unit slope
shown in Fig. \ref{fig:sfrco}, and the dotted one to a robust regression
as described in the text: log(SFR/\lco)\,=\,$(-2.25\pm0.15)\,$[\logoh]\,$+ (11.31\pm1.3)$.
}
\label{fig:sfrco_oh}
\end{figure}

Judging from our sample alone,
the offset between SFRs and high \lco\ and low \lco\
is apparently independent of metallicity, at least to the lowest oxygen
abundance of our sample, \logoh$\sim$7.7.
However, \cgcg\ at this metallicity is only tentatively detected, and 
our other detected galaxies have a relatively narrow spread in metallicities, \logoh$\sim$8.
If we include in the regression the \citet{schruba12} compilation and their five \cotwo\ detections (converted to \coone\
as described above), we find a significant correlation 
between the SFR/\lco\ ratio and \logoh, as shown in Fig. \ref{fig:sfrco_oh}
(Pearson parametric correlation coefficient, $-$0.88,
$>$99.999\% significance level). 
Two galaxies are discrepant with this metallicity trend, \cgcg\ and \sbs.
\sbs\ was not directly observed in \coone, so there may be some problems with the
multiplicative factor relating CO(3--2) to CO(1--0) \citep[see][]{hunt14}.
\cgcg, on the other hand, is directly observed in \coone\ but 
the detection is only tentative. %falls more than 1$\sigma$ below the expected trend.
More CO(1--0) observations at metallicities below \logoh\,=\,8 are needed to
verify this inconsistency, especially because of the expectation that
for \zzsun$\la$0.1, CO may cease to be a viable tracer of \htwo\ \citep{bolatto13}.
We discuss the implications of the relation between SFR/\lco\ and abundance in the following section.

\section{Molecular depletion times, the CO-to-\htwo\ conversion factor, and specific SFR}
\label{sec:depletion}

Although we have shown that \lco\ is correlated with SFR even at low metallicity,
the main parameter of interest is \htwo\ mass, \mhtwo.
In fact,
the study of low-metallicity star formation and \htwo\ content is hampered by the unknown 
factor, \aco, used to convert observed CO intensities to \mhtwo.
Observationally, \mhtwo, \aco, and the 
time required to consume the molecular gas reservoir, 
\tdep, 
are intimately related and difficult to disentangle.
By definition,
\tdep\ depends inversely on the star-formation rate 
\tdep\,$\equiv$\,\mhtwo/SFR\,=\,\lco\,\aco/SFR, but
the potential simplicity of predicting \mhtwo\ from SFR 
(or vice versa) through \tdep\ is compromised by the necessary CO-to-\htwo\ factor.
%\aco: \mhtwo/SFR\,=\,\aco\ \lco/SFR\,=\,\tdep.
The observed (or directly derivable) quantities are \lco\ and SFR; thus only the ratio, \tdep/\aco, is constrained.

Much effort has been devoted to finding the dependence of \aco\ on metal abundance
\citep[e.g.,][]{wilson95,israel97,rosolowsky03,bolatto08,leroy09b,leroy11,genzel12,schruba12,bolatto13}.
High-resolution cloud virial mass measurements tend to find little or no dependence 
on metallicity \citep{wilson95,rosolowsky03,bolatto08}.
However, because CO is found only in dense regions, rather than in
diffuse gas, \aco\ depends on the physical scales probed by the observations,
especially at low metallicity \citep{rubio93}. 
Given the relatively constant \tdep\ in nearby star-forming spirals
\citep[e.g.,][]{bigiel08,leroy08,bigiel11},
some groups have assumed
a constant \tdep\ to infer \aco\ in metal-poor galaxies.
Such an assumption tends to give a steep dependence of \aco\ on O/H: 
\aco$^{-1}\propto$\,(\zzsun)$^{\sim 2-3}$ \citep{genzel12,schruba12}. 
An alternative approach relies on IR observations to 
derive \mhtwo\ and infer \aco\ by estimating the dust-to-gas ratio and assuming
it is the same in adjacent regions with/without \htwo. 
This method has the considerable advantage of being independent of CO observations altogether,
but requires comparable quantities of both \hi\ and \htwo, and relies on the assumption that 
emissivity properties of dust grains are the same in dense and diffuse environments.
Recent work using this method, introduced by \citet{israel97},
also gives super-linear slopes, although
somewhat shallower (and with large scatter) than those obtained with the assumption of 
constant \tdep: 
\aco$^{-1}\propto$\,(\zzsun)$^{\sim 1.6-2}$ \citep{gratier10,leroy11,bolatto11}. 

\begin{figure}[!h]
\includegraphics[angle=0,width=\linewidth]{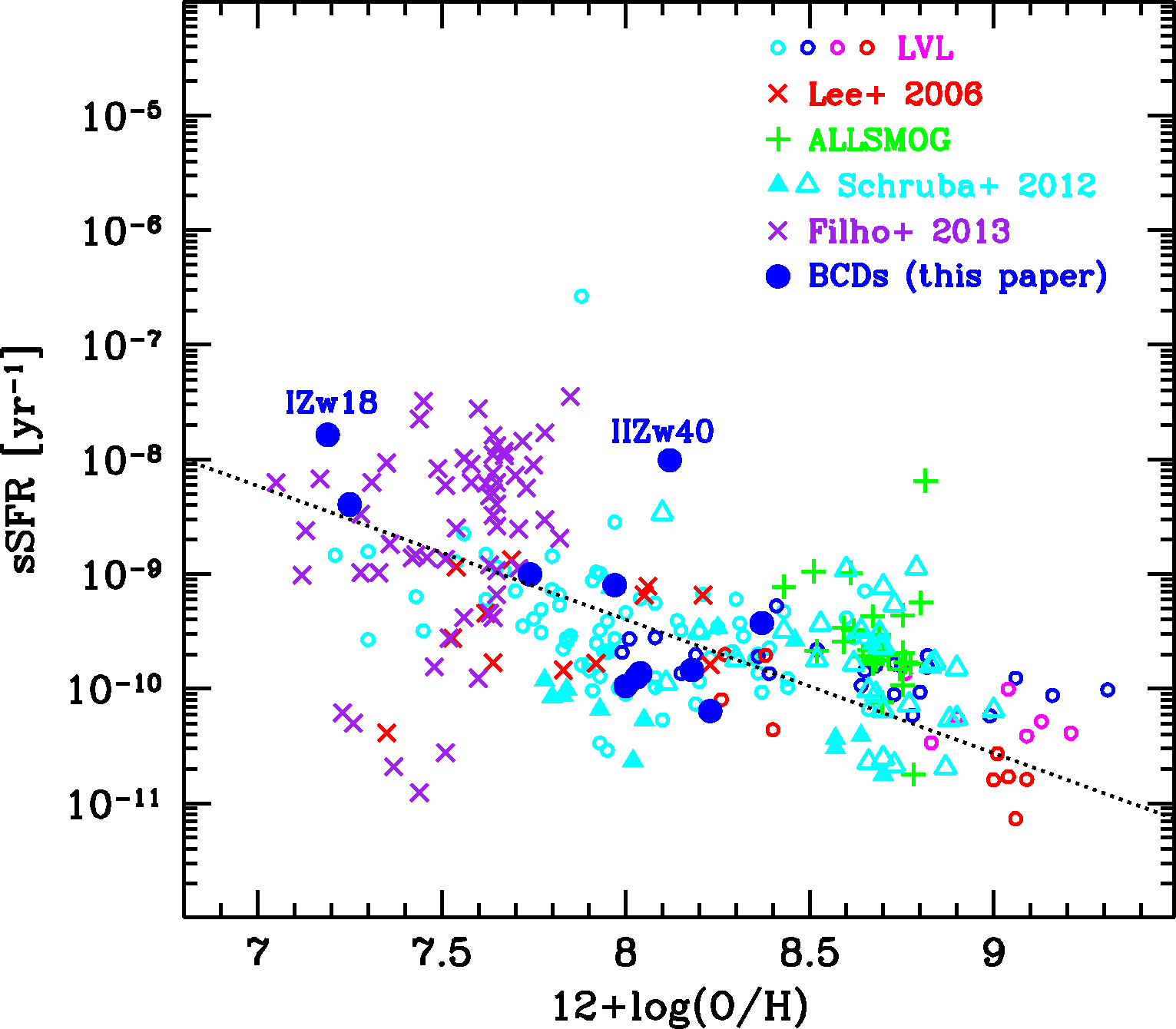}
\caption{Specific SFR vs. \logoh\ for our sample and including data from
the literature \citep{hunt12,schruba12,filho13,bothwell14}.
The metallicities for the ALLSMOG \citep{bothwell14} sample have been modified
as discussed in the text.
As in Fig. \ref{fig:sfrco},
the open triangles for \citet{schruba12} correspond to their compilation of data from
the literature and the filled ones to their metal-poor dwarf observations.
The regression line is a robust best-fit of the Local Volume Legacy \citep[LVL,][]{lee09} 
and the dwarf irregulars from \citet{lee06}.
The various colors for the LVL give the Hubble type (early types red, magenta; late types blue, cyan).
}
\label{fig:ssfroh}
\end{figure}

Here we present a different approach to constrain \aco\ at low metallicity.
We exploit the results by \citet{saintonge11b} \citep[see also][]{huang14}
who find a significant inverse correlation of \tdep\ with sSFR:
%specific SFR (sSFR\,=\,SFR/\mstar):
galaxies with higher sSFR also have shorter molecular gas depletion times.
However, sSFR tends to also be inversely correlated with metallicity
\citep[e.g.,][]{salim14}, as shown in 
Fig. \ref{fig:ssfroh} \citep[adapted from][]{hunt12}. 
Thus, we must find a way to disentangle the effects of sSFR 
on \tdep\ from those of metallicity.

\subsection{The CO-to-\htwo\ conversion factor \label{sec:aco}}

\citet{saintonge11b} found that star-forming galaxies of lower mass have
much shorter \tdep\ than more massive ones, while the depletion time for
atomic gas remains relatively constant.
They also found that \tdep\ is strongly correlated with sSFR,
with a single relation being able to fit both ``normal'' \citep[``main-sequence'', e.g.,][]{salim07}
galaxies and more extreme starbursts including LIRGs and ULIRGs.
Their relation is based on a single value of \aco, appropriate for the 
properties of their sample, selected by stellar mass.

\begin{figure}[ht!]
%\includegraphics[angle=0,width=0.965\linewidth]{LCOSFRvsSSFRALLSMOG}
% 14/9/2015 reply to resubmission
\includegraphics[angle=0,width=0.965\linewidth]{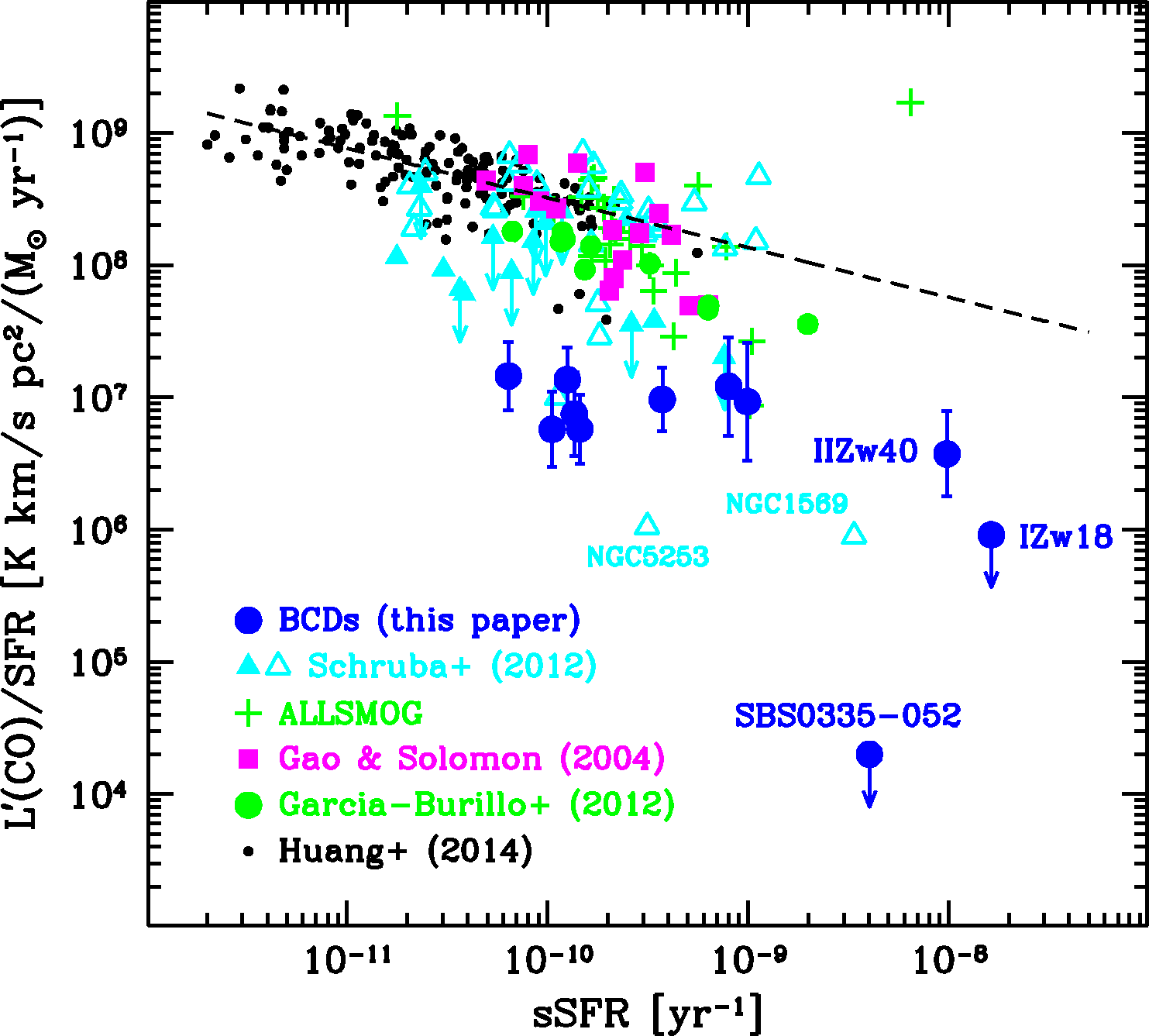}
\caption{\lco/SFR\ vs. sSFR for our previous IRAM sample and including
%data from the literature \citep[COLDGASS:][]{saintonge11b,gao04,garciaburillo12,schruba12,hunt14}.
data from the literature \citep[COLDGASS:][]{huang14,gao04,garciaburillo12,schruba12,hunt14}.
Our best-fit of the COLDGASS$+$Gao\&Solomon$+$Garcia-Burillo samples is shown (which 
is consistent with the regression given by \citet{saintonge11b} for only the COLDGASS galaxies).
\label{fig:lcosfr}
}
\end{figure}

Here we explore the hypothesis that the Saintonge et al. correlation between \tdep\ and sSFR
extends to the lower-mass (and higher sSFR) galaxies in our sample.
However, for a better comparison with our SFR estimates, we use
the COLDGASS data by \citet{huang14} who re-evaluated the SFRs within the IRAM beam
using additional mid-infrared and UV data for COLDGASS galaxies. 
Because there is some evidence that the SFRs from fitting of spectral-energy
distributions such as done by \citet{saintonge11b} are overestimated relative to those 
from \ha$+$24\,\micron\ or FUV$+$24\,\micron\ \citep[e.g.,][]{leroy13a,huang14},
the comparison with the COLDGASS sample should be more consistent using the
SFRs by \citet{huang14}\footnote{We have performed the entire analysis with
both datasets and confirm an overestimate of $\sim$0.2\,dex in the SFRs
measured by \citet{saintonge11b} relative to \citet{huang14}.
This change affects only slightly the metallicity dependence of \aco, steepening 
the power-law index by 0.08 
%{\bf 
with respect to the \citet{huang14} data.}. %}}.
In addition to the COLDGASS data,
we also consider other data from the literature where both CO measurements and
stellar masses are available (see Sect. \ref{sec:comparison}).
The dashed line in Fig. \ref{fig:lcosfr} indicates our fit with 
a robust linear regression to COLDGASS data 
\citep[using the SFRs and \mstar\
within a 22\arcsec\ aperture by][]{huang14},
together with the samples of \citet{gao04} and \citet{garciaburillo12}: 
log(\lco/SFR)\,=\,($-0.38\pm0.03)\,$log(sSFR)\,$+(\,4.75\pm0.31)$.
The new fit is consistent with that found by \citet{huang14} for only 
the COLDGASS galaxies.  
However, both the dwarf galaxies in our sample and those in 
\citet{schruba12} show large residuals relative to the general trend.

\begin{figure*}[!htbp]
\vspace{0pt}
\hbox{
\centerline{
\includegraphics[angle=0,height=0.38\linewidth]{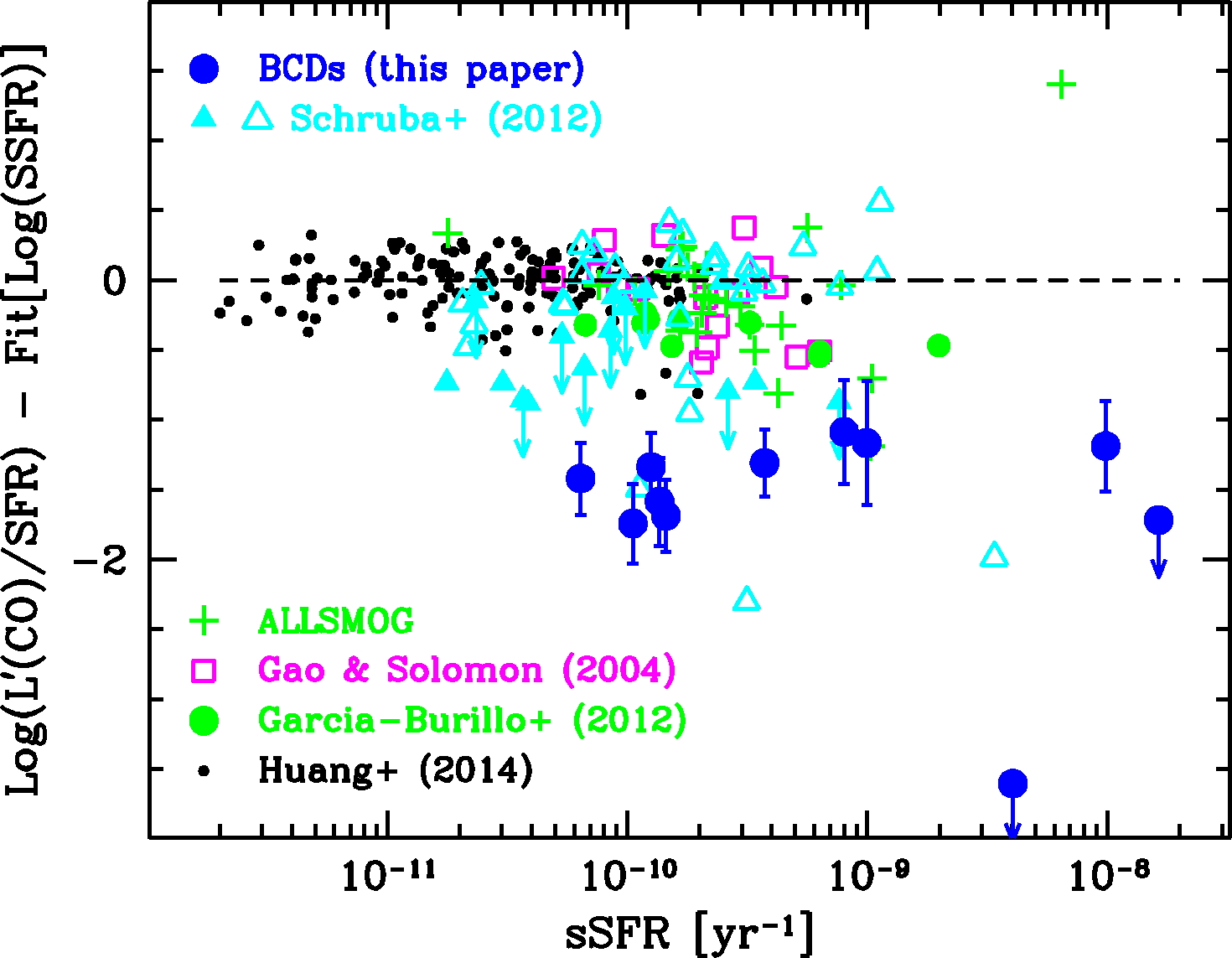} 
\hspace{-0.2\baselineskip}
\includegraphics[angle=0,height=0.38\linewidth]{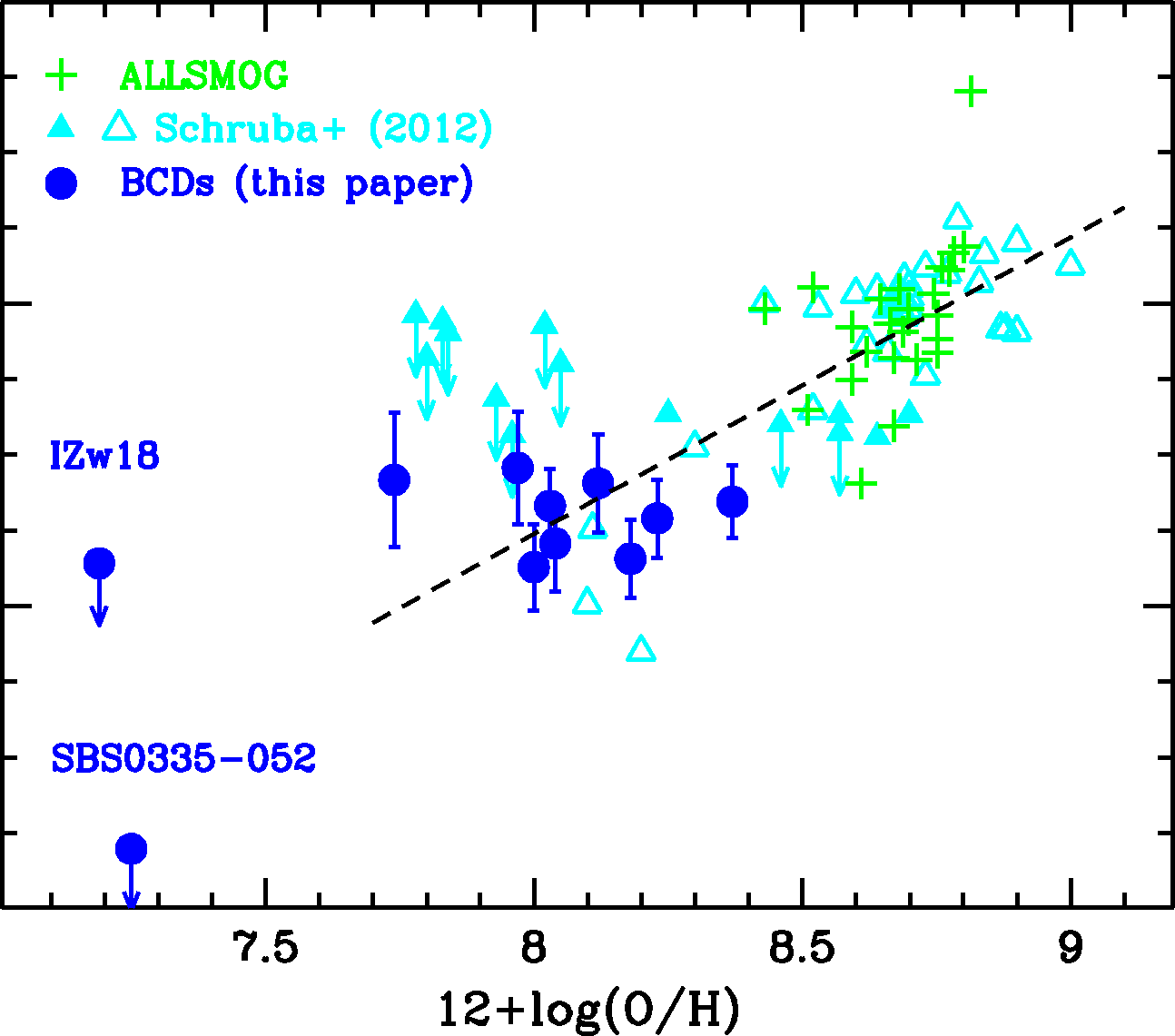} 
}
}
%\vspace{-1.5\baselineskip}
\caption{\label{fig:cosfr_res}
Residuals from the regression shown in Fig. \ref{fig:lcosfr} vs. sSFR (left panel)
and vs. \logoh\ (right).
We include here data for our sample and 
literature data \citep[COLDGASS:][]{huang14,gao04,garciaburillo12,schruba12,bothwell14}.
The horizontal dashed line in the left panel shows 0 to aid the eye,
while the regression in the right panel is the robust fit to the significant
trend of the residuals of the \lco/SFR fit with metallicity $\propto$(\zzsun)$^{1.96}$.
%NGC\,1140 and NGC\,1156 [\logoh$\sim$8.2] are both significantly larger than the IRAM 2.8\,mm beam
%for CO(1-0) and are thus apparently missing CO flux (SFR values are global ones).
}
\end{figure*}

The residuals to this fit are plotted vs. sSFR and O/H in Fig. \ref{fig:cosfr_res}.
Because \tdep\,=\,\aco\ \lco/SFR, 
the residuals from the correlation of \lco/SFR on sSFR 
should eliminate the dependence on \tdep:
[~\lco/SFR~](observed)/\tdep (fit)\,=\,[~\aco(Z)/$\alpha_{\rm CO}(\zsun)~]^{-1}$.
Thus, these residuals should correspond directly to the inverse of \aco, relative to \zsun.
The left panel of Fig. \ref{fig:cosfr_res} demonstrates that there is no additional
correlation with sSFR not accounted for by the COLDGASS relation;
however, the right panel shows that there is a clear correlation with O/H. 
The correlation of the residuals with metallicity is able to almost completely rectify the
deviation of the low-metallicity galaxies by almost two orders of magnitude 
from the COLDGASS galaxies.
We find a robust fitted slope of \aco$^{-1}\propto$\,(\zzsun)$^{1.96\pm0.16}$, 
more shallow than found with the assumption of constant \tdep\ \citep{schruba12,genzel12},
but consistent with the \aco\ metallicity dependence found
by using dust continuum measurements \citep{leroy11,bolatto11}.

\begin{figure*}[!htbp]
\hbox{
\centerline{
\includegraphics[angle=0,height=0.38\linewidth]{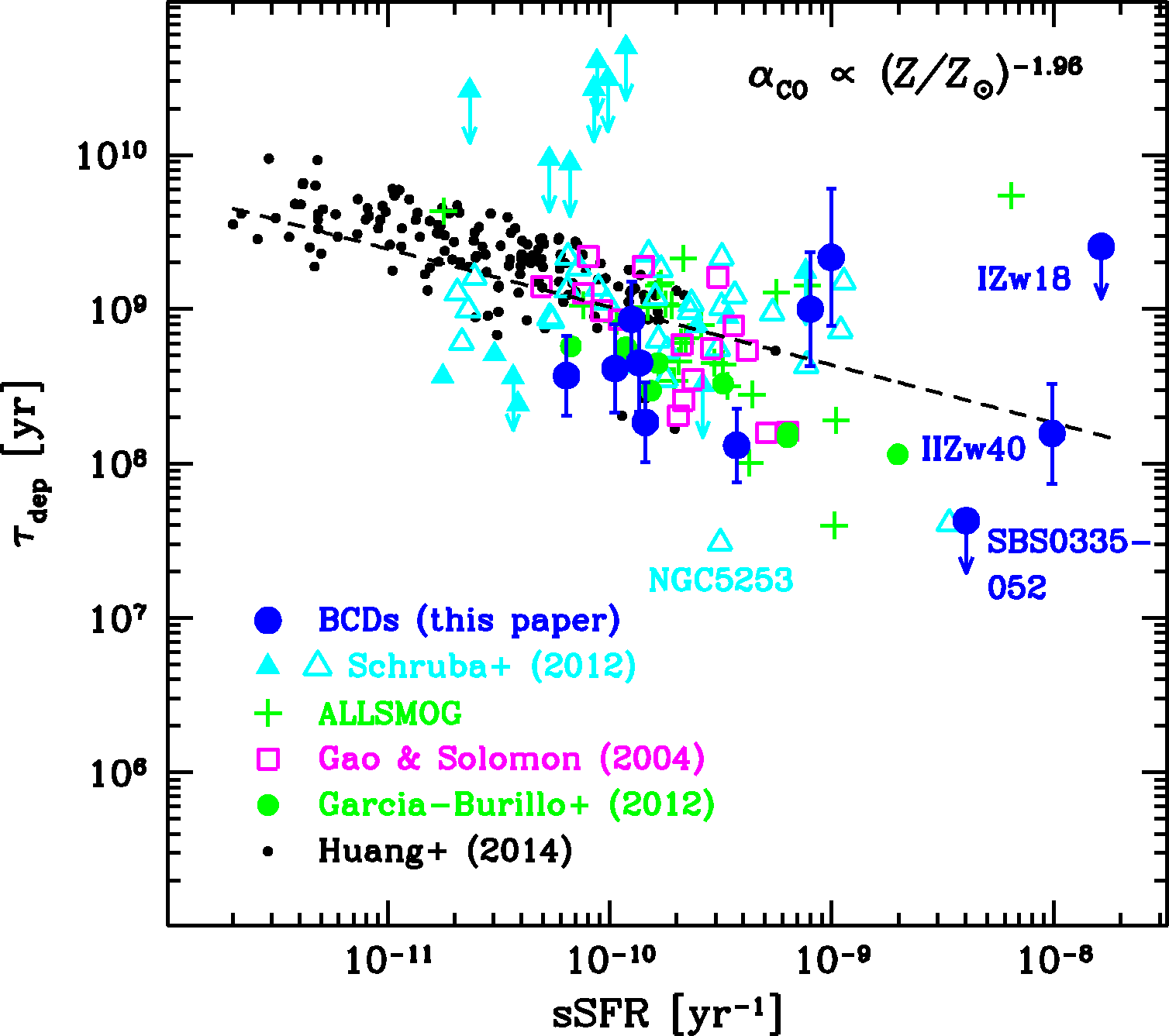}
\hspace{-0.2\baselineskip}
\includegraphics[angle=0,height=0.38\linewidth]{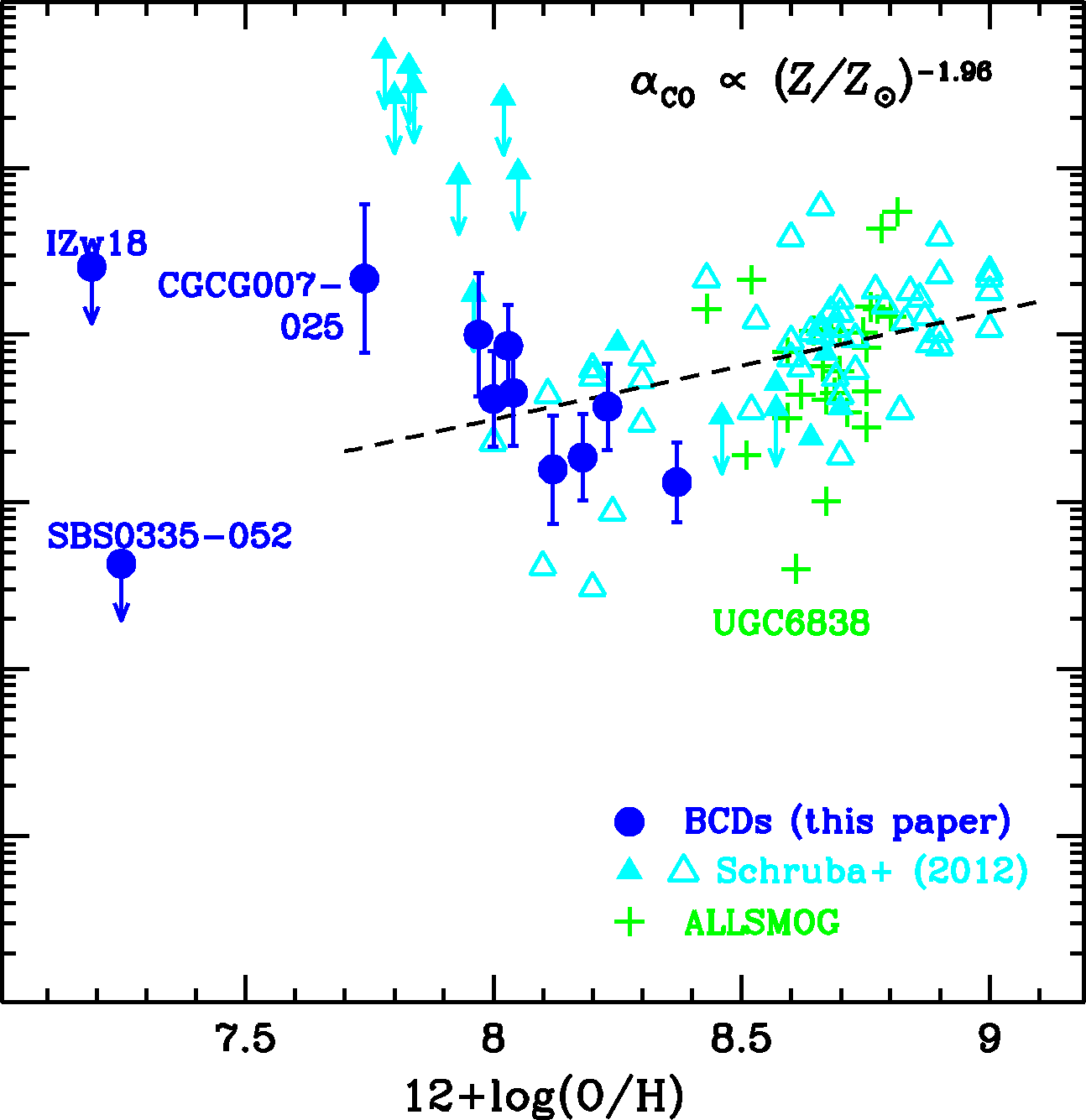}
}
}
\caption{\tdep\ vs. sSFR (left panel) and vs. \logoh\ (right)
for our sample and including 
data from the literature \citep[COLDGASS:][]{huang14,gao04,garciaburillo12,schruba12,bothwell14}.
As in previous figures,
the open triangles for \citet{schruba12} correspond to their compilation of data from
the literature and the filled ones to their metal-poor dwarf observations.
The regression line is our fit to the \citet{schruba12} and our sample
as described in the text.
\label{fig:tdep}
}
\end{figure*}

For the calculation of \mhtwo, we adopt a solar value \aco\,=\,3.2\,\msun\,(\kkms\,pc$^{2}$)$^{-1}$, 
not including a factor of 1.36 for helium \citep[see also][]{saintonge11b}.
For our galaxies and those from \citet{schruba12} and \citet{bothwell14} having
metallicities \zzsun $<$1 (\logoh$<$8.69),
we applied \aco$\propto$\,(\zzsun)$^{-1.96}$;
for metallicities \zzsun$\geq$1, we used the constant \aco\ as given above. 
Figure \ref{fig:tdep} shows the result of converting our \lco\ observations to \mhtwo.
Although there is some scatter, 
% and a few galaxies may be missing CO flux
%(e.g., NGC\,1156 and NGC\,1140 because of their large apparent size relative
%to the 22\arcsec\ IRAM beam), 
the trend of \tdep\ with sSFR established by \citet{saintonge11b,huang14} 
seems to extend to the higher sSFRs and shorter \tdep\ of these dwarf galaxies.
A steeper dependence of \aco\ on metallicity would be inconsistent with this trend,
but more data at \logoh$\la$8 are needed to better assess any systematics.

In any case, in agreement with previous work \citep[e.g.,][]{saintonge11b,bothwell14,huang14},
our analysis shows a strong variation of the molecular gas depletion times, with
values ranging from $\sim$10\,Gyr for the COLDGASS galaxies
to $\sim$70\,Myr for II\,Zw\,40 and even smaller, $\la$10\,Myr for \sbs,
NGC\,5253 and NGC\,1569 \citep[from][]{schruba12}.
This variation is correlated with a range in sSFRs of almost three orders of magnitude.

The right panel of Fig. \ref{fig:tdep} suggests that \tdep\ depends weakly on metallicity,
even after the \aco\ dependence has been accounted for.
The regression line is the robust fit to our galaxies and those from \citet{schruba12};
the correlation is significant (Pearson coefficient of 0.38, significance $>$99.9\%),
with a slope of 0.64$\pm$0.11.
A similar (weak) effect was also found by \citet{leroy13a}, but over a narrower range in metallicities
and \tdep.
Because \tdep\ and sSFR are correlated, and sSFR and O/H are also correlated (see Fig. \ref{fig:ssfroh}), 
such a result is perhaps not surprising; 
it could be the result of the secondary dependence on sSFR or on \mstar.
In Sect. \ref{sec:discussion}, we discuss
forms of \aco\ and its metallicity dependence
\citep[e.g.,][]{wolfire10,krumholz11}. 

\begin{figure}[!htbp]
\includegraphics[angle=0,width=0.965\linewidth]{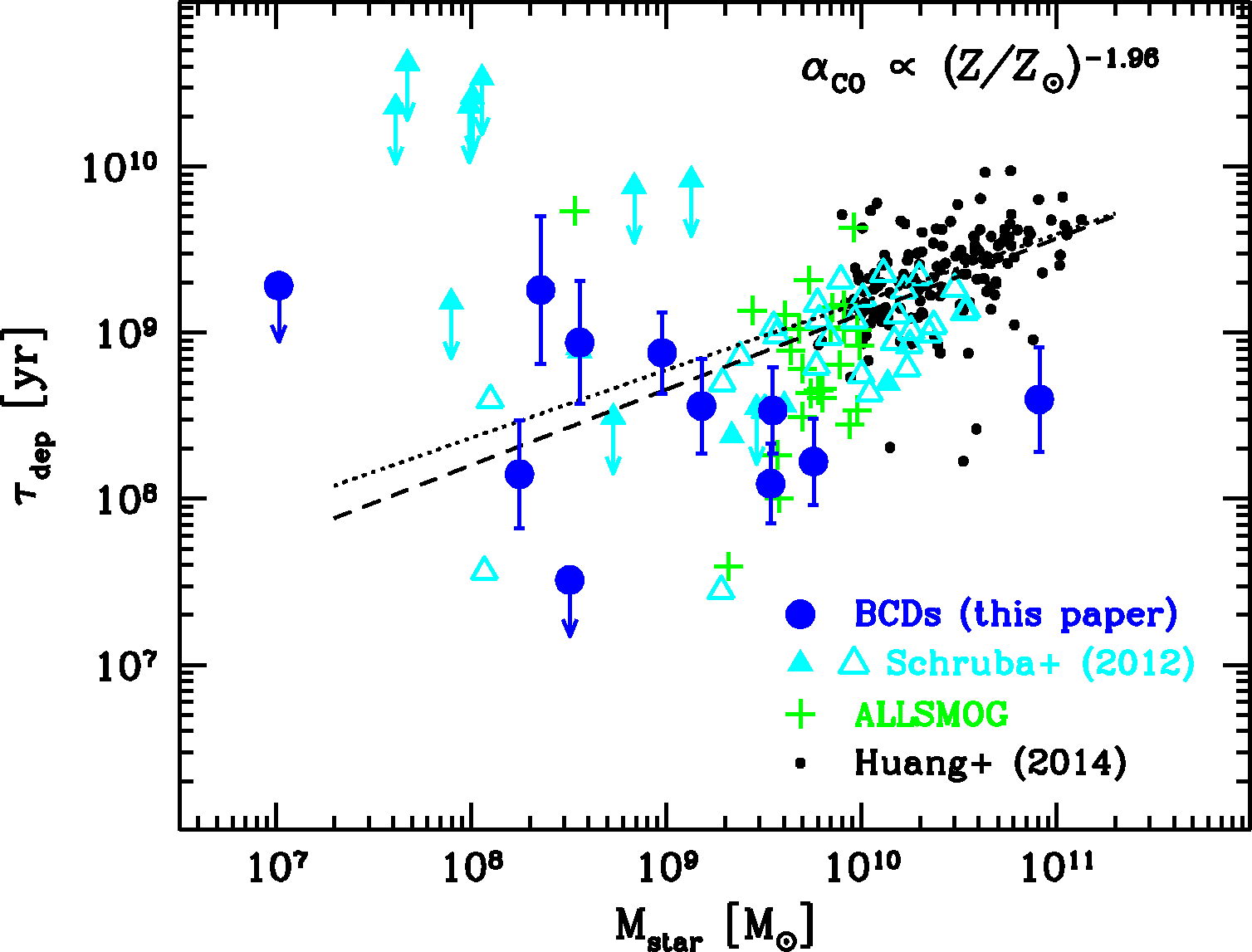}
\caption{\tdep\ vs. \mstar\ 
for our sample and including data from
%the literature \citep{saintonge11b,schruba12,bothwell14}.
the literature \citep{huang14,schruba12,bothwell14}.
As in previous figures,
the open triangles for \citet{schruba12} correspond to their compilation of data from
the literature and the filled ones to their metal-poor dwarf observations.
Our robust best-fit regression for these samples is shown as a dashed line;
%the slope found by \citet{saintonge11b} for the COLDGASS galaxies alone is
the dotted line corresponds to the slope for the COLDGASS galaxies alone \citep{huang14}. 
\label{fig:tdepmstar}
}
\end{figure}

Because O/H and \mstar\ are correlated through the mass-metallicity relation,
in Fig. \ref{fig:tdepmstar}, we have investigated the possibility that \tdep\ depends on \mstar. 
This correlation is also formally significant,
even more so than the one with O/H (Pearson coefficient of 0.56, significance $>$99.999\%).
The slope of 0.46$\pm$0.03 (shown by a dashed line in Fig. \ref{fig:tdepmstar}) is similar 
to that (0.41$\pm$0.05) found for the COLDGASS galaxies alone. 
Our results suggest that
low-mass metal-poor galaxies tend to have high sSFR (Fig. \ref{fig:ssfroh})
and shorter \htwo\ depletion times (Figs. \ref{fig:tdep}, \ref{fig:tdepmstar}), and the two properties are correlated.
This result is not altogether new, although we have shown that it extends to much
lower stellar masses and higher sSFRs than the COLDGASS sample.
%However, this conclusion may be sample dependent because the galaxies studied here are bona fide
%starbursts, while other low-mass metal-poor galaxies particularly in the Local Group may be more 
%quiescent with longer gas depletion times \citep[e.g.,][]{schruba12}.

\section{Atomic and molecular gas content}
\label{sec:atomic}

The implication is that for low-mass, 
metal-poor galaxies, with high sSFR, \tdep\ is systematically shorter than for more massive, metal-rich systems.
%probably because of higher gas-mass fractions at high sSFR and low metallicity. 
Because our sample combined with the \citet{schruba12} and the \citet{filho13} compilation spans 
a wide range of O/H and sSFR, with \coone\ detections down to \zzsun$\sim$0.1,
it is well suited to
assess the dependence on these parameters of molecular and total gas (\hi$+$\htwo)
content, relative to stellar mass and total baryonic mass. 
The \hi\ masses, \mhi, for our sample are given in Table \ref{tab:sample} (see also
Sect. \ref{sec:hi}) and we have taken \mhi\ for the \citet{schruba12} galaxies
from \citet[THINGS,][]{walter08} when available, or otherwise from \citet{kennicutt03}.

\begin{figure*}[!htbp]
\hbox{
\centerline{
\includegraphics[angle=0,height=0.38\linewidth]{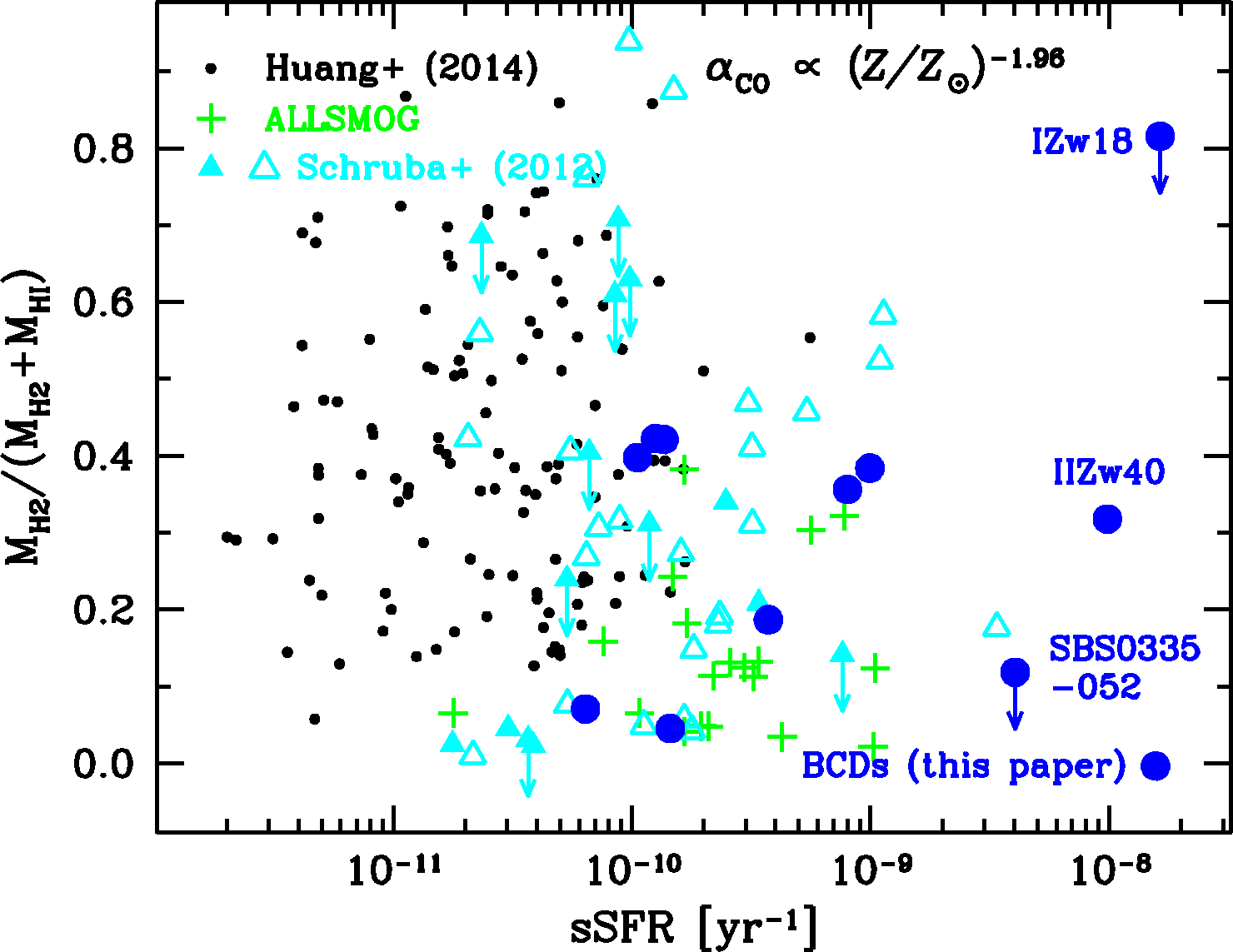}
\hspace{-0.2\baselineskip}
\includegraphics[angle=0,height=0.38\linewidth]{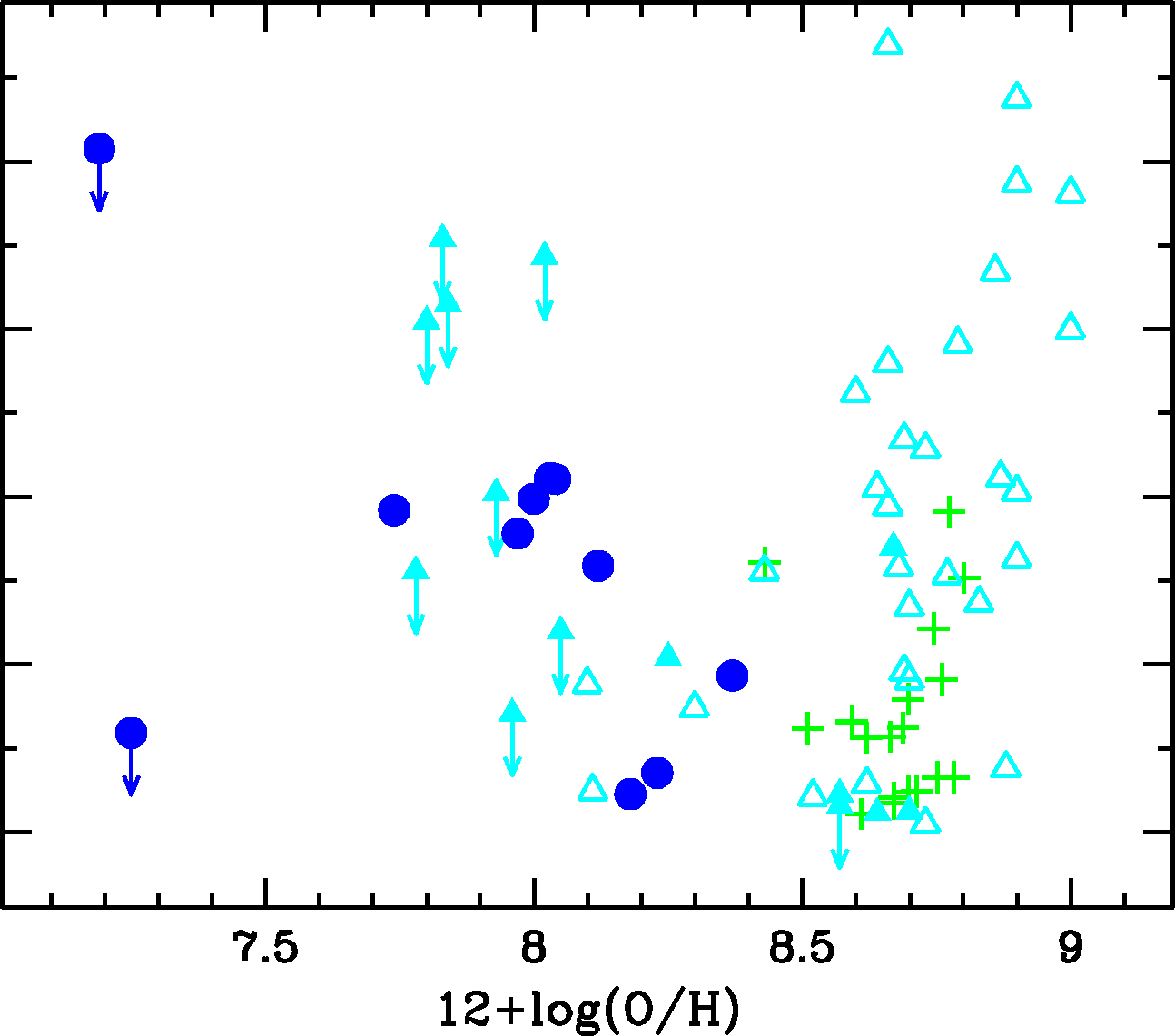}
}
}
\caption{\mhtwo/M$_{\rm gas}$\,=\,\mhtwo/(\mhi\ $+$ \mhtwo) vs. sSFR (left panel) and vs. \logoh\ (right)
for our sample and including data from
the literature \citep{schruba12,bothwell14};
%the left panel gives also the COLD GASS detections \citep{saintonge11b}.
the left panel gives also the COLDGASS data \citep{huang14}. 
As in previous figures,
the open triangles for \citet{schruba12} correspond to their compilation of data from
the literature and the filled ones to their metal-poor dwarf observations.
}
\label{fig:h2frac}
\end{figure*}

We first investigate any trends of \mhtwo-to-total gas mass fraction in
Fig. \ref{fig:h2frac}, both as a function of sSFR (left panel) and \logoh\ (right).
%To better compare \htwo\ and \hi\ content in the BCDs, most of which have dimensions
%comparable to the IRAM beam,
%in Fig. \ref{fig:h2frac} we use the \hi\ mass within the galaxy when available
%(Col. 11 in Table \ref{tab:sample}).
We use total \hi\ mass for this comparison, although it is almost certainly 
more extended than the \htwo\ (see Table \ref{tab:sample}), which usually resides within
the optical confines of the galaxy
\citep{regan01,leroy09a}.
%\footnote{We have used here
%the \hi\ mass only within the galaxy when available (Col. 11 in Table \ref{tab:sample}).}.
%While for the galaxies in our sample, there could be a weak trend for higher \htwo\
%fractions to be associated with higher sSFR, the combined samples show no
%correlation.
%The same is true for trends with O/H, although \iizw\ with a very high
%\htwo\ fraction is not particularly metal poor.
The combined samples suggest
a weak trend for \htwo\ fractions to decrease at low sSFR, but there is no trend with O/H. 
A better assessment of both trends would require more CO data from high-sSFR and metal-poor galaxies
(and resolved \hi\ measurements).
%The \htwo\ fraction never exceeds $\sim$20\% in our low-metallicity sample.
%There is apparently no correlation with \htwo\ fraction, either with sSFR or O/H, 
%probably because of the difficulty of the comparison.
%We have used total \hi\ masses from the
%literature, while the \htwo\ content is inferred from the 22\arcsec\ IRAM beam.
%It would have been much more appropriate to compare gas masses within the same beam,
%but this was possible only for a small fraction of our sample (see Sect. \ref{sec:hi}).

\begin{figure}[!htbp]
\includegraphics[angle=0,width=0.965\linewidth]{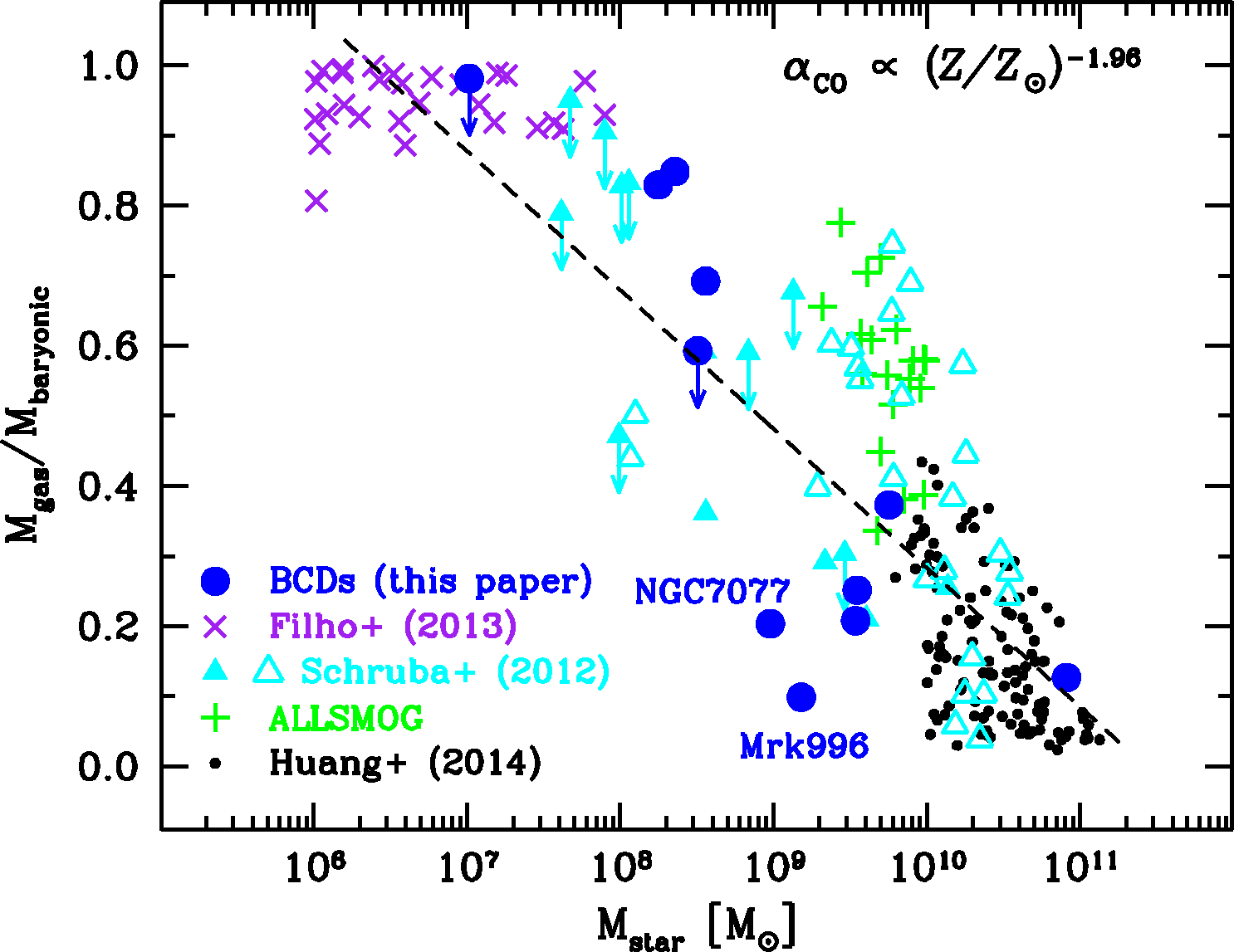}
\caption{\mgas\,=\,(\mhi\ $+$ \mhtwo)/\mbary\ vs. \mstar, including
data from the literature \citep[COLDGASS:][]{huang14,schruba12,filho13,bothwell14}.
Our robust best-fit regression for these samples is shown as a dashed line.
\label{fig:gasfrac_mstar}
}
\end{figure}

Fig. \ref{fig:gasfrac_mstar} shows the total gas \mgas\ ($\equiv$\,\mhtwo$+$\mhi) fraction relative
to the total baryonic mass \mbary\ ($\equiv$\,\mgas$+$\mstar) as a function of stellar mass. 
%To better compare with stellar mass, and the \hi\ contribution to the baryonic mass budget,
%we adopt here and below} 
As in Fig. \ref{fig:h2frac}, we use 
the total \hi\ mass for our targets (Col. 12 in Table \ref{tab:sample}).
By including also the data from \citet{schruba12,filho13,bothwell14,huang14},
we find there is a significant correlation between gas mass fraction and \mstar, as shown
by the robust regression line.
This is mainly due to the spread in \mstar\ of the two most extreme samples,
COLDGASS and \citet{filho13}.
At low gas-mass fractions, this may also be partially induced by the co-dependence of the two axes, 
because of the presence of \mstar.
Nevertheless,
%although the low-mass end of the correlation is dominated by the XMP sample \citep{filho13}, 
%for which stellar masses may be underestimated (see Sect. \ref{sec:comparison}), 
the trend is roughly continuous over a factor of $10^5$ of \mstar, from $\sim 10^6$ to $\sim 10^{11}$\,\msun.

\begin{figure*}[!htbp]
\hbox{
\centerline{
\includegraphics[angle=0,height=0.38\linewidth]{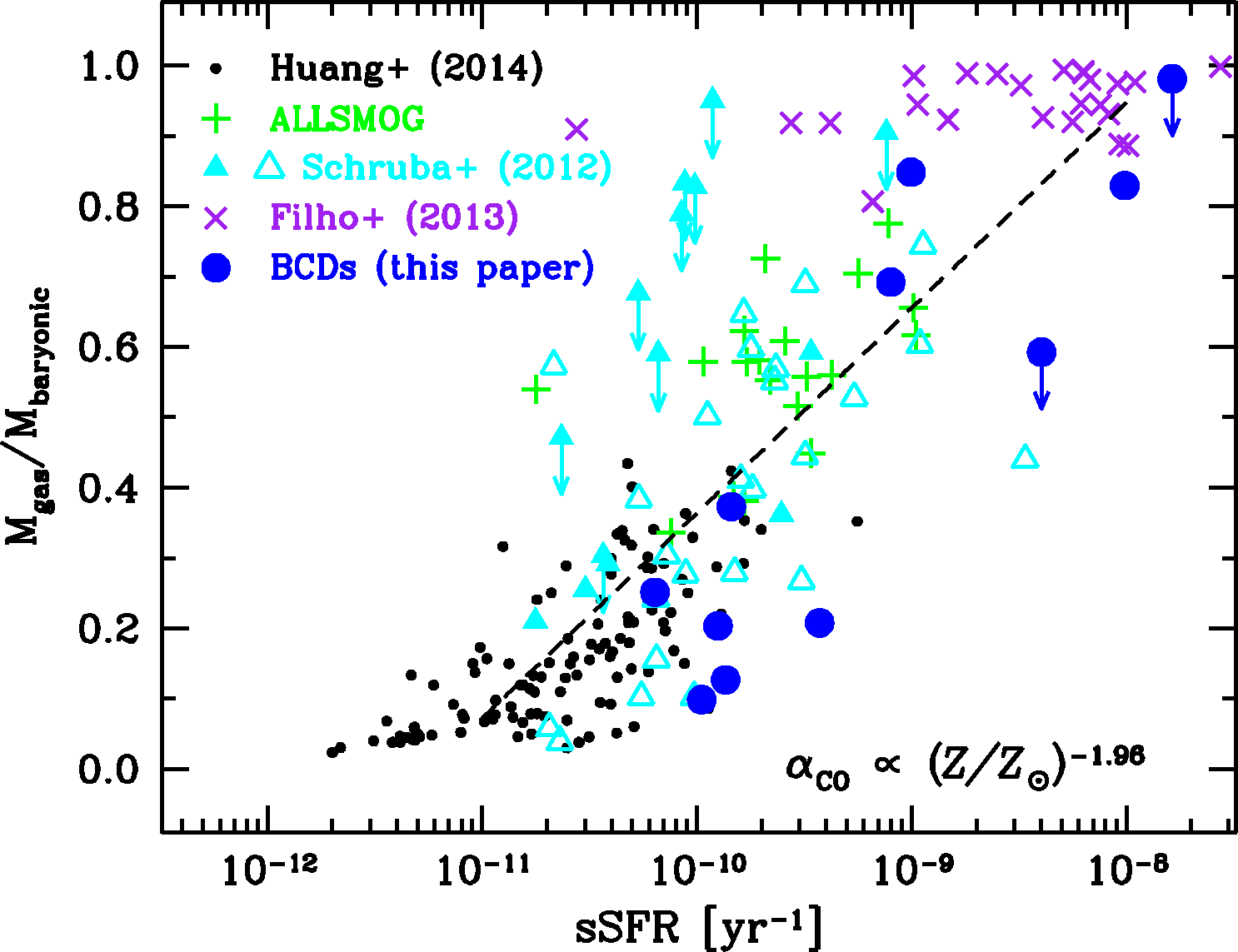}
\hspace{-0.2\baselineskip}
\includegraphics[angle=0,height=0.38\linewidth]{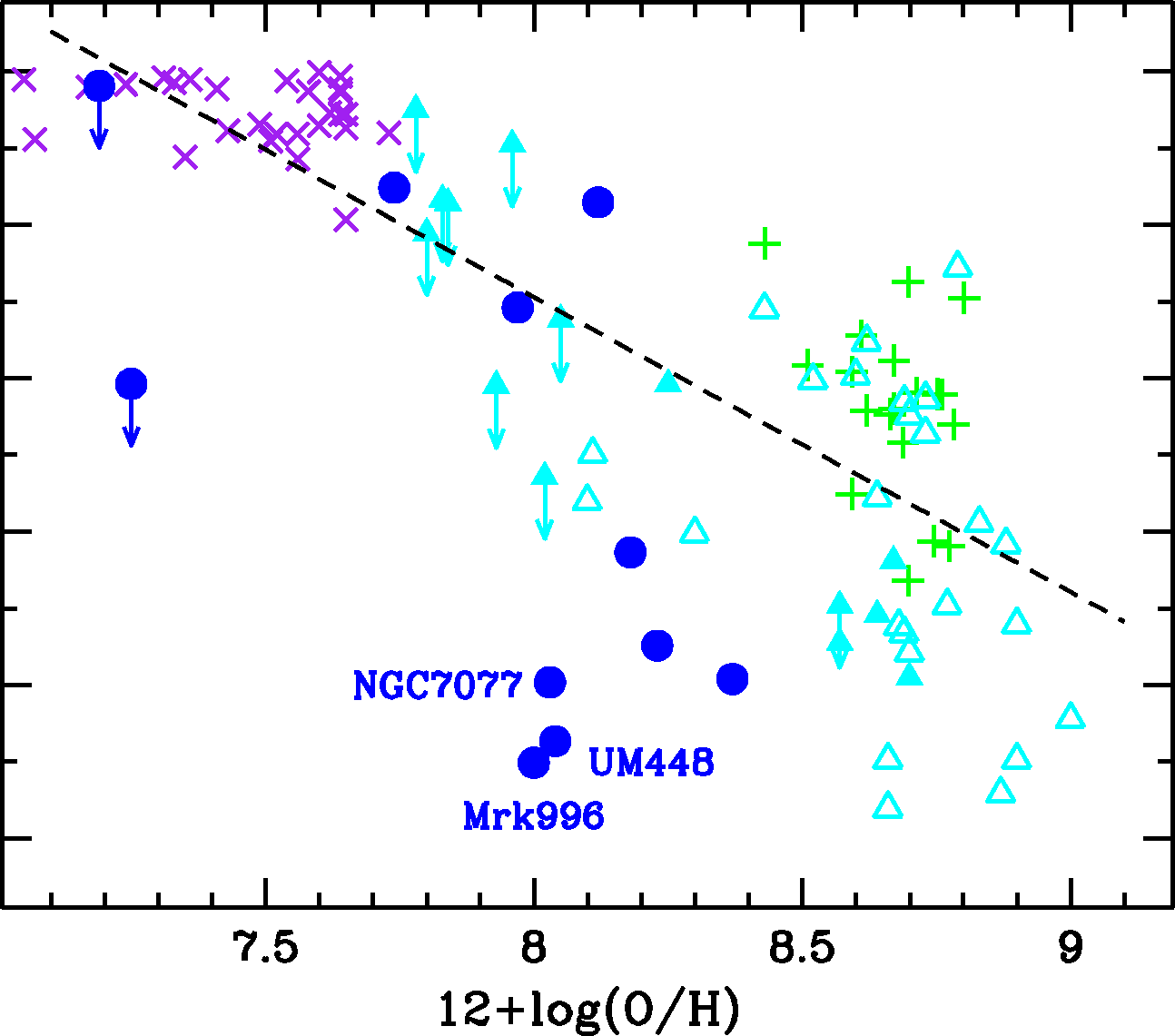}
}
}
\caption{\mgas\,=\,(\mhi\ $+$ \mhtwo)/\mbary\ vs. sSFR (left panel) and vs. \logoh\ (right)
for our sample and including data from
the literature \citep{schruba12,filho13,bothwell14};
the left panel gives also the COLDGASS detections \citep{saintonge11b}.
As in previous figures,
the open triangles for \citet{schruba12} correspond to their compilation of data from
the literature and the filled ones to their metal-poor dwarf observations.
The regression lines shown in both panels correspond to our fit of the galaxies from 
\citet{schruba12,filho13}, ALLSMOG, and our sample as described in the text
(with all the uncertainties alluded to in Sect. \ref{sec:comparison}).
}
\label{fig:gasfrac}
\end{figure*}

\begin{figure*}[!htbp]
\hbox{
\centerline{
\includegraphics[angle=0,height=0.38\linewidth]{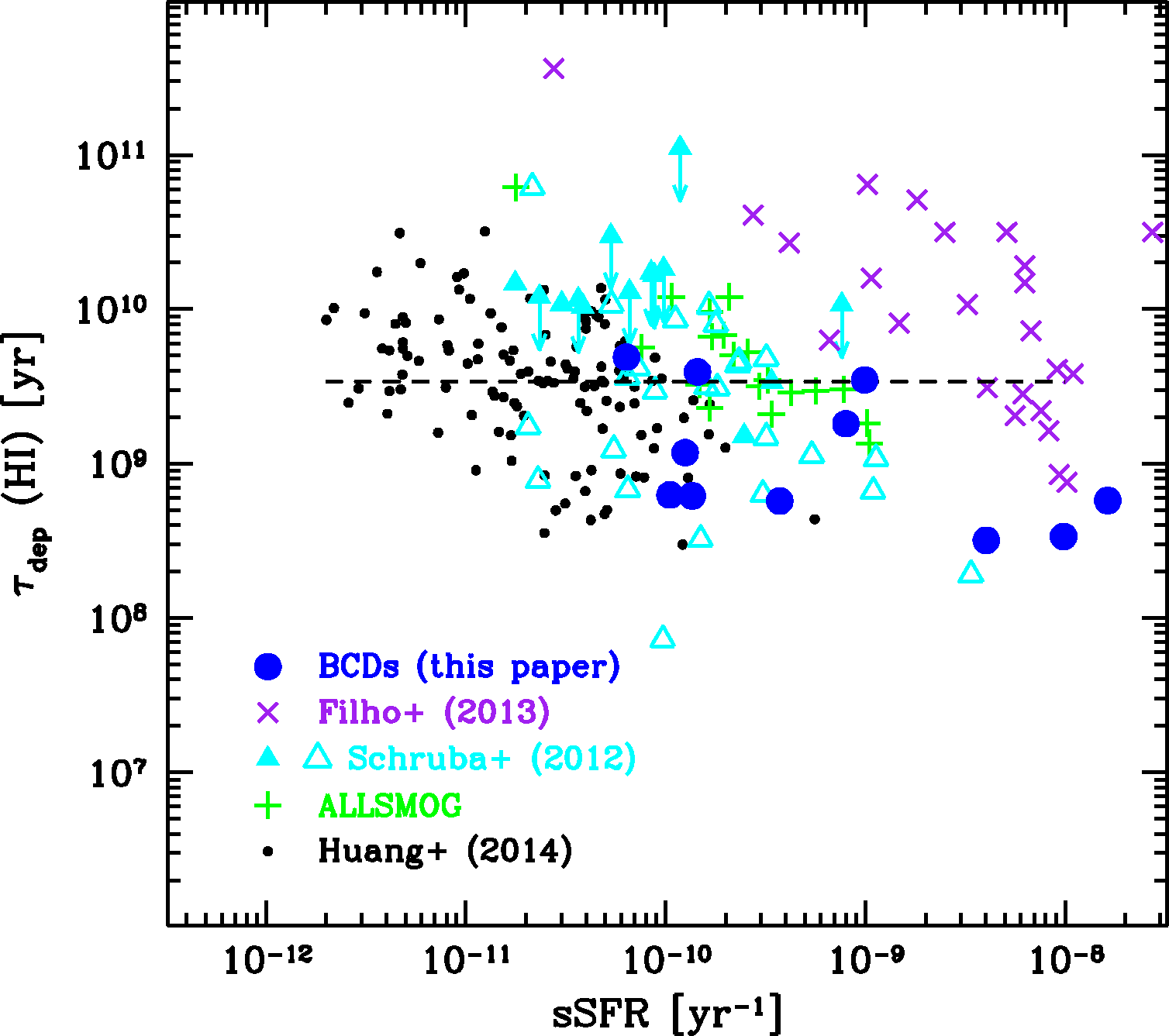}
\hspace{-0.2\baselineskip}
\includegraphics[angle=0,height=0.38\linewidth]{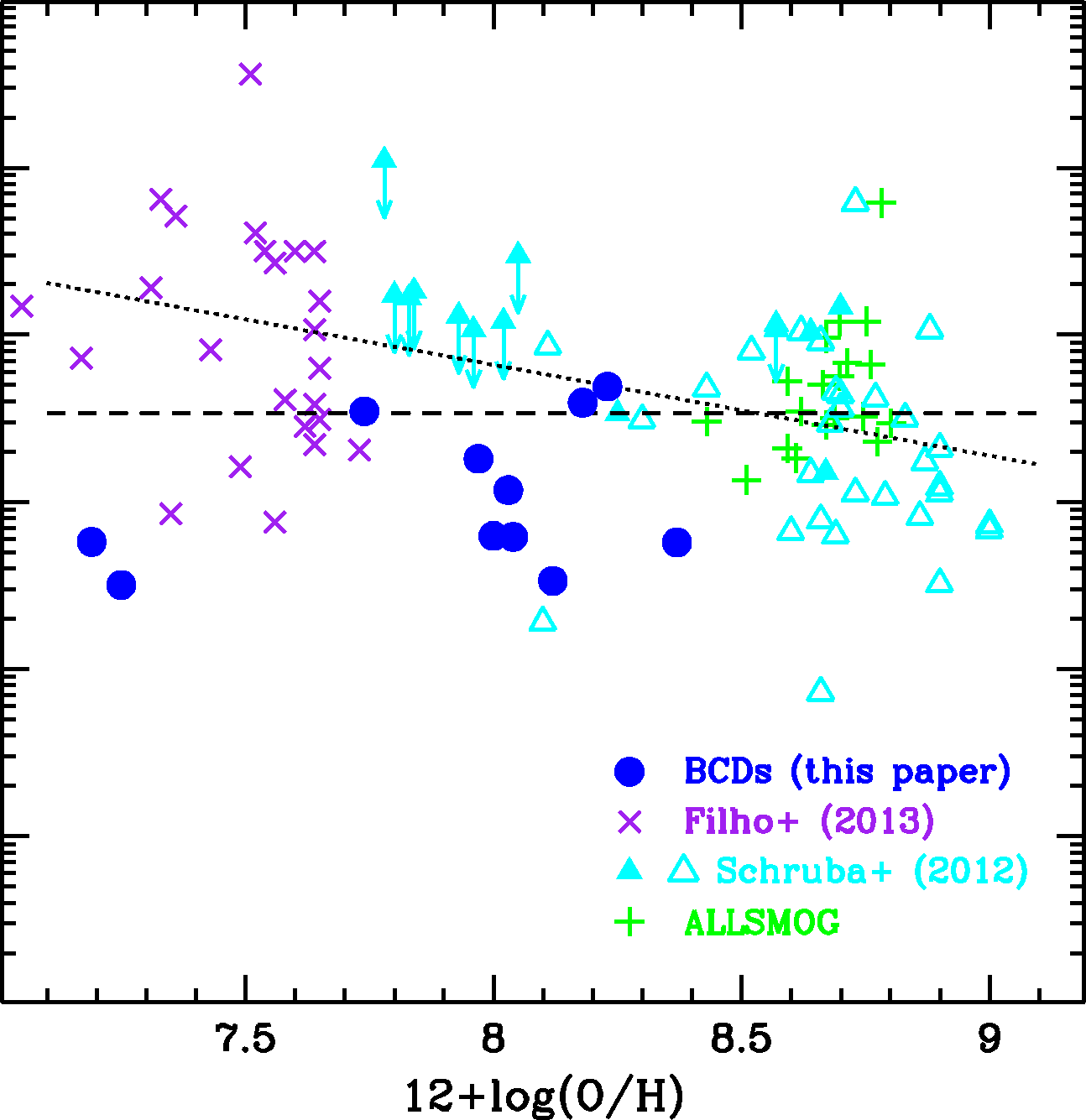}
}
}
\caption{\tdep (\hi) vs. sSFR (left panel) and vs. \logoh\ (right)
for our sample and including data from
the literature \citep{schruba12,bothwell14}.
As in previous figures,
the open triangles for \citet{schruba12} correspond to their compilation of data from
the literature and the filled ones to their metal-poor dwarf observations.
The constant \tdep (\hi), shown by a horizontal dashed line,
corresponds to 3.4\,Gyr as found for the GASS sample
\citep{schiminovich10}.
}
\label{fig:tdephi}
\end{figure*}

Two of our sample galaxies are clear outliers, Mrk\,996 and NGC\,7077,
both of which have a smooth elliptical-type morphology despite their low
metallicity.
Thus they may have already exhausted much of their \hi, since the \htwo-to-total gas mass
fractions are relatively high, $\sim$40\%. 
These two galaxies could be in the process of quenching their star formation,
unless more gas is accreted from the outside. 

We examine in Fig. \ref{fig:gasfrac}
the trend of total gas mass relative to \mbary\ vs.
sSFR (left panel) and \logoh\ (right). 
Here also the total gas fraction is significantly correlated with
both sSFR and \logoh\ (see the robust regression lines), although there are
significant outliers in our sample for the metallicity trend, which
is mainly dominated by the XMP galaxies.
However, as in Fig. \ref{fig:gasfrac_mstar}, there may be a co-dependence of
the two axes because of the presence of \mstar\ (left panel) and because
of the stellar mass-metallicity correlation (right).

The \hi\ depletion times, \tdep (\hi), for our combined sample are
shown in Fig. \ref{fig:tdephi}.
There is more than an order-of-magnitude scatter, but no systematic
trend relative to sSFR.
This result is similar to that found for more massive galaxies,
and the horizontal dashed line
in Fig. \ref{fig:tdephi} corresponds to the roughly constant
\tdep (\hi) of 3.4\,Gyr found for the GASS galaxies by \citet{schiminovich10}. 
Fig. \ref{fig:tdephi} shows a possible weak trend with O/H,
which starts to affect \tdep (\hi) at \logoh$\la$8.
This could be a consequence of the large fraction of \hi-dominated
galaxies at these metallicities \citep[e.g.,][]{krumholz13}. 

In general, we find that galaxies are more gas rich at low stellar masses,
at high sSFRs, and at low metallicities, although the metallicity
trends tend to have more scatter.
Similar conclusions have been reached by 
\citet[][]{bothwell14} and by \citet{filho13}, although for a 
more limited range in parameter space.
Our combined samples span five orders of magnitude in \mstar,
a factor of $\sim 10^{3}$ in sSFR, and
a factor of $\sim$100 in oxygen abundance.
The trends for total gas fraction are similar to those we find for the molecular
gas depletion time \tdep: low-mass, metal-poor, and high sSFR galaxies tend to have
shorter \tdep. 
\tdep\ decreases by a factor of $\sim$100 over a variation of $\sim$1000 in sSFR
and a factor of $\sim$10 in O/H.
However, \hi\ depletion times are roughly constant, although with large scatter, across this 
entire range of sSFR and O/H.

\section{Discussion}
\label{sec:discussion}

The short molecular gas depletion times, \tdep$\la$100\,Myr, we find at the highest
sSFRs ($\ga$0.1 Gyr$^{-1}$) are comparable to starbursts at $z\ga1$; 
they are smaller than those found for ``main-sequence'' galaxies at
similar redshifts \citep{genzel15}.
Thus, the low-metallicity galaxies studied here may be low-mass analogues
of starbursts observed at cosmological distances.
This is not unlikely since most of them are likely merging or interacting
systems (see Sect. \ref{sec:source}).
A starburst could enhance star-formation efficiency,
namely the quantity of \htwo\ converted into stars over a free-fall
time \citep[e.g.,][]{krumholz09}; mergers or interactions would be expected
to drive gas into the central regions, causing higher gas densities which
could, at least partly, compensate for the low metallicity through increased
self shielding. 

Independently of the origin of the short molecular gas depletion times,
the secondary dependence of \tdep\ on \mstar\ and/or metallicity is not straightforward
to interpret.
Because of the way we have derived the molecular gas depletion times, 
the determination of a power-law dependence of \aco\ on metallicity is intimately related to \tdep.
To better understand whether short \tdep\ and its dependence on \mstar\
and O/H are artefacts of our derivation of \aco,  
here we investigate an alternative formulation of \aco,
and its potential relation with \tdep.
What we have defined as \tdep\ is the inverse of what
\citet{krumholz11} call specific gas SFRs,
SSFR$_{\rm gas}$(CO,\htwo,total),
the ratio of SFR surface densities
to the various gas components of the ISM: 
\sigmagas, \sigmahtwo, and \sigmahi. 
Given a total gas mass surface density, \sigmagas,
Krumholz et al. give precise predictions for the fraction of 
molecular gas mass in 
the form of \htwo, \fhtwo, as well as the fraction of
molecular gas where CO dominates over other forms of carbon, \fco. 

In their models \citep[see also][]{krumholz09,mckee10}, \fhtwo\
is a function of the ratio of the interstellar radiation field (ISRF)
at the surface of the gas cloud, \gnot, and the hydrogen number volume density, $n_H$,
as well as the dust optical depth to UV photons, \taudust.
\fhtwo\ depends on metallicity through \taudust\ which 
is assumed to be linearly proportional to \zzsun.
The transition in a cloud of atomic and ionized carbon to CO is driven
mainly by dust extinction, thus cloud optical depth. 
The fraction of gas mass where the carbon is in CO form, \fco,
depends on \fhtwo, modified by an exponential function which depends
only weakly on \gnot, and $n_H$, but strongly on metallicity through
\av\ (i.e., \taudust), assumed to vary linearly with \zzsun\ 
\citep[for more details see][]{wolfire10}.
In these models,
SFR surface density depends on \fhtwo\ and \sigmagas, and 
is defined by a roughly constant timescale, 2.6\,Gyr.
Thus, by assuming a characteristic value for \gnot/$n_H$, \citet{krumholz11}
derive the dependence of the gas SSFRs 
on gas mass surface density, \sigmagas, and metallicity, \zzsun. 

\begin{figure}[!htbp]
\includegraphics[angle=0,width=0.965\linewidth]{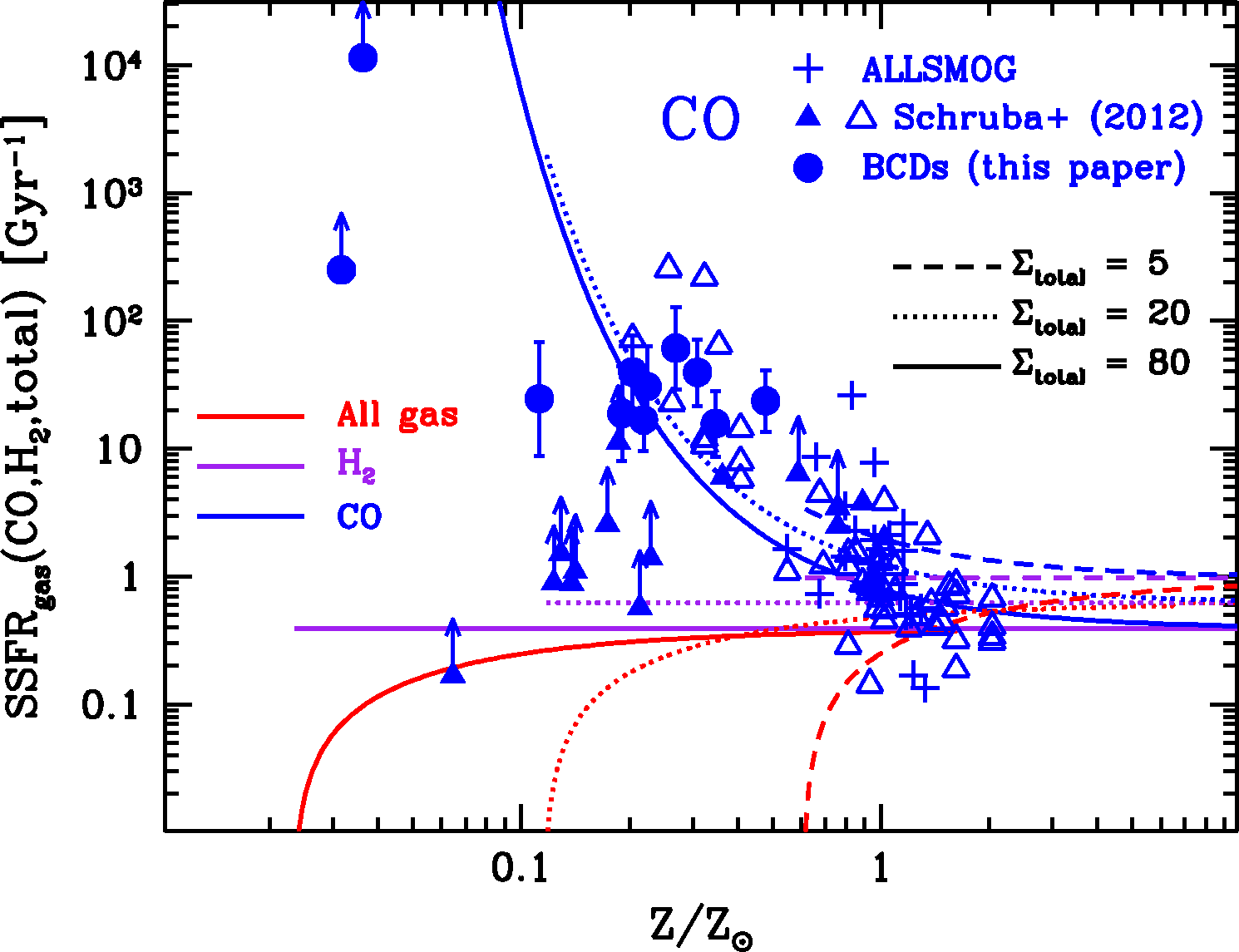}
\caption{Ratio of SFR and CO-inferred \htwo\ mass vs. relative metallicity as described in the text, 
including our data and data from the literature \citep[COLDGASS:][]{schruba12,bothwell14}.
As in previous figures,
the open triangles for \citet{schruba12} correspond to their compilation of data from
the literature and the filled ones to their metal-poor dwarf observations.
Following \citet{krumholz11}, here \mhtwo\ is inferred with a constant 
\aco\,=\,4.4\,\msun\,(\kkms\,pc$^{2}$)$^{-1}$, as described in the text. 
Also shown are the predictions by \citet{krumholz11}, showing 
three values for total gas surface density, $\Sigma_{\rm gas}$:
5\,\msunpc, 20\,\msunpc, and 80\,\msunpc\ indicated by
dashed, dotted, and solid lines, respectively.
Total gas corresponds to red, purple to \htwo, and blue to CO.
\label{fig:krumholz_co}
}
\end{figure}

Fig. \ref{fig:krumholz_co} shows their predictions together with the data
we have analyzed here.
Although the models are formulated in terms of surface densities,
to better compare with global values \citet{krumholz11}
include a resolution correction factor ($c$), which we have adopted
according to their prescription.
Following \citet{krumholz11},
the data points correspond to molecular gas masses, \mhtwo(CO), inferred from CO
luminosities using a constant conversion factor,
\aco\,=\,4.4\,\msun\,(\kkms\,pc$^{2}$)$^{-1}$; this is the same \aco\ as 
used up to now \citep[see][]{saintonge11b} but here with the inclusion of helium.
The CO-inferred molecular mass has been derived assuming a total gas surface density 
\sigmagas\,=\,20\,\msunpc;
as shown by the blue curves in Fig. \ref{fig:krumholz_co}, the
exact value of \sigmagas\ is not critical.
As emphasized by \citet{krumholz11}, this inferred mass \mhtwo(CO) should 
not be interpreted as the total mass of molecular gas, but rather the 
mass of the molecular gas where the predominant form of carbon is CO.
Also shown in Fig. \ref{fig:krumholz_co} are the \citet{krumholz11} model 
predictions\footnote{Here we continue to use
the 8.69 solar calibration of \citet{asplund09} rather than the value
of \logoh\,=\,8.79 used by \citet{krumholz11}.}
as a function of metallicity, \zzsun,
for three values of total gas column density
\citep[for more details see][]{krumholz11}. 
The models of CO-inferred \htwo\ mass differ from the true \htwo\ mass by a
multiplicative factor that varies exponentially with \zzsun\
\citep[because of the form of \fco\ described above, see also][]{wolfire10,bolatto13}. 
As in \citet{krumholz11}, the SSFR$_{\rm gas}$(\htwo, CO) curves are shown only for 
metallicities and total gas surface densities that
are sufficiently high that the \htwo\ and CO mass fractions are non-zero.
Like the data compilation presented by \citet{krumholz11},
our data are also qualitatively consistent with the trends predicted by the models.

\begin{figure*}[!htbp]
\hbox{
\centerline{
\includegraphics[angle=0,height=0.38\linewidth]{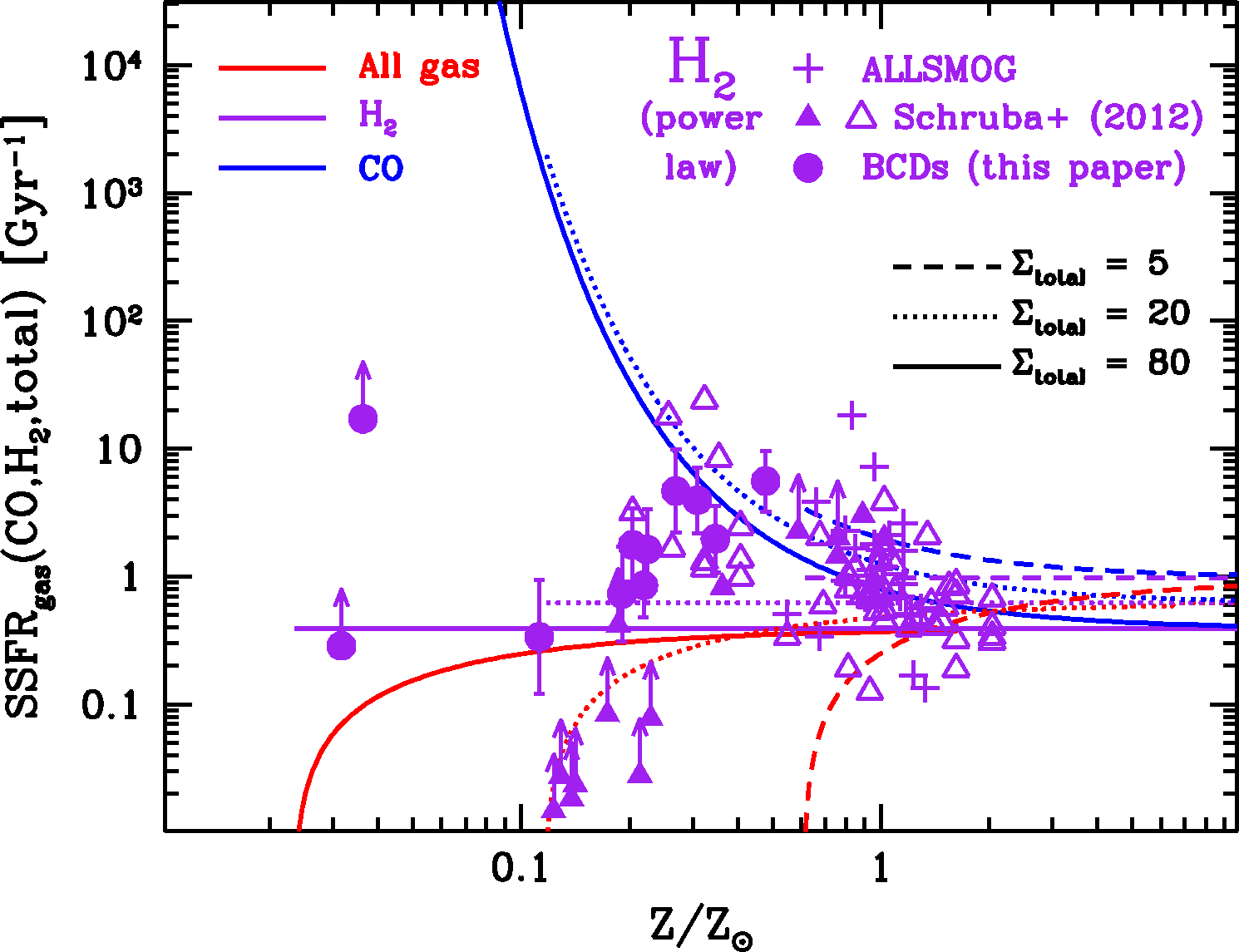}
\hspace{-0.2\baselineskip}
\includegraphics[angle=0,height=0.38\linewidth]{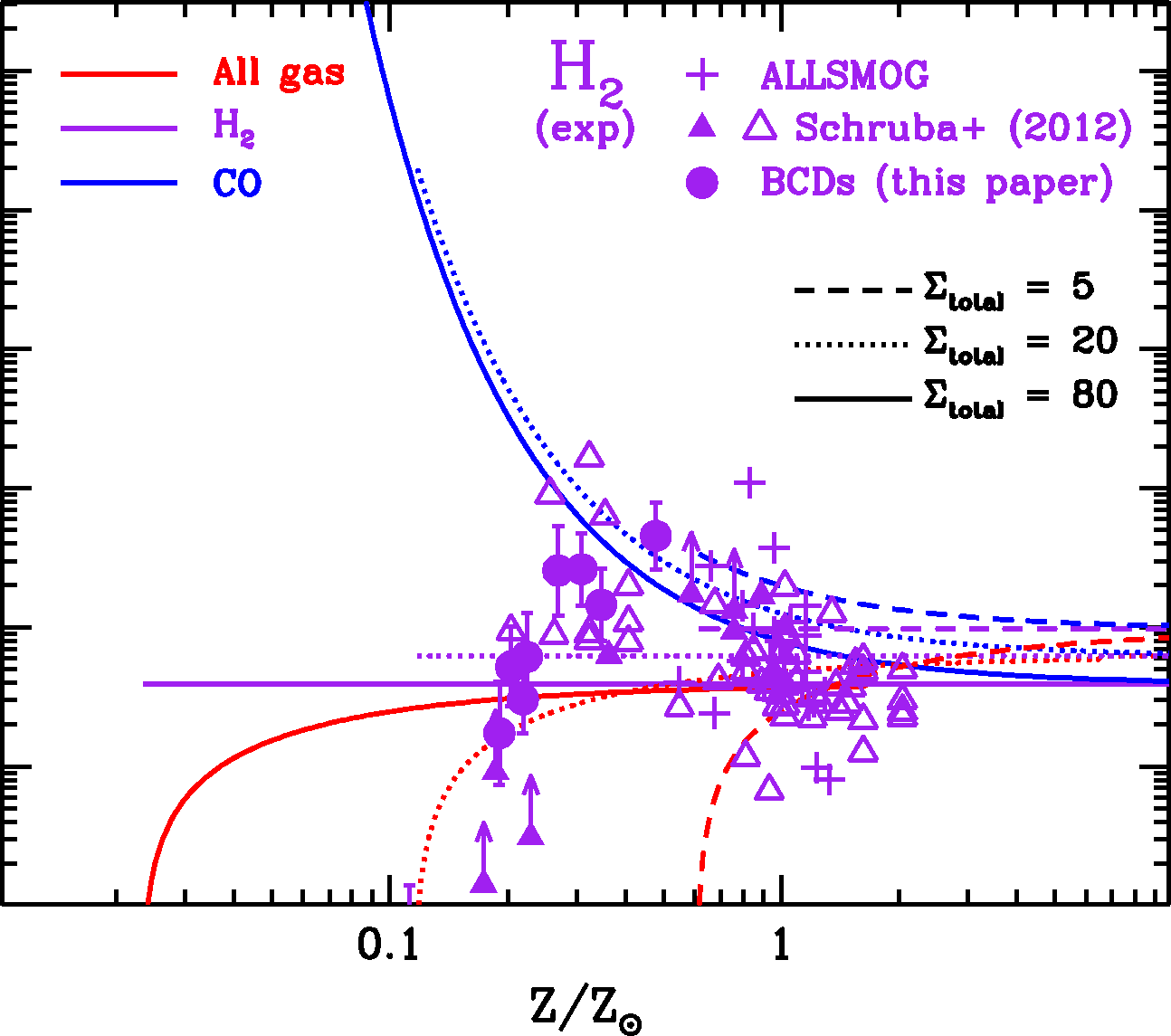}
}
}
\caption{Ratio of SFR and CO-inferred \htwo\ mass vs. relative metallicity as described in the text, 
including our data and data from the literature \citep[COLDGASS:][]{huang14,schruba12,bothwell14}.
As in previous figures,
the open triangles for \citet{schruba12} correspond to their compilation of data from
the literature and the filled ones to their metal-poor dwarf observations.
In the left panel, \mhtwo\ is inferred with our formulation of \aco\ as a power-law dependence
on O/H as described in the text;
in the right, we use the exponentially-dependent \aco\ as found by \citet{wolfire10,krumholz11}. 
Also shown are the predictions by \citet{krumholz11}, showing 
three values for total gas surface density, $\Sigma_{\rm gas}$:
5\,\msunpc, 20\,\msunpc, and 80\,\msunpc\ indicated by
dashed, dotted, and solid lines, respectively.
Red color shows total gas, purple \htwo, and blue CO.
\label{fig:krumholz_h2}
}
\end{figure*}

Figure \ref{fig:krumholz_co} shows that
the data and the models agree extremely well for the prediction of CO luminosity;
however at significantly sub-solar metallicities, \zzsun$\la$0.2, 
there are some discrepancies relative to the \htwo\ prediction, with
some data exceeding by a factor of
3 or more the expected variation of the \aco\ factor.
% and the lowest metallicity detection, \cgcg\ ($\sim$0.1\,\zzsun),
%falls below the prediction by more than a factor of 10.
This is shown in Fig. \ref{fig:krumholz_h2} where 
we compare the power-law dependence of \aco\ derived previously
with the exponential variation predicted by the models.
Unlike in Fig. \ref{fig:krumholz_co} where \mhtwo(CO) depended on
a constant \aco, here we derive \mhtwo\ from a metallicity-dependent \aco;
the left panel shows the power-law dependency \aco\,$\propto$\zzsun$^{-1.9}$
as shown in previous figures, and the right panel the exponential
dependency found by \citet{wolfire10} and \citet{krumholz11}, \aco\,$\propto\exp(-$\zzsun$)$.
As discussed above, such a dependence arises
through the assumption that \aco\ depends exponentially on the visual extinction
within the clouds, \av, and that \av\ is proportional to \zzsun.

Because both \mhtwo\ and SFR predicted by the
\citet{krumholz11} models depend on \fhtwo, the expectation is that
SSFR(\htwo) should be constant, independently of metallicity.
Fig. \ref{fig:krumholz_h2} shows that the data do not quantitatively follow this
trend, but rather show a residual increase of SSFR(\htwo) toward
low metallicities.
Neither formulation (power-law or exponential) of \aco\ gives a horizontal
trend of SSFR(\htwo) as a function of \zzsun. 
The exponential \aco\ seems to better approximate the models around solar
metallicity, but overestimates \aco\ %by a factor of 10 or so 
at \zzsun$\sim$0.1.
On the other hand, the power-law \aco\ reproduces well the lowest metallicity
SSFR(\htwo) (perhaps by construction, given that our \aco\ was heavily influenced
by the lowest metallicity galaxy in our sample).
%However, the metallicities \zzsun$\la$0.2 are farther
%from the constant trend of SSFR(\htwo). 
In any case, it seems that the short depletion times, corresponding to high
SSFR(gas) are not a result of our approach, nor is our power-law formulation
of \aco\ significantly inferior to the exponential one.

As shown in Fig. \ref{fig:krumholz_h2},
galaxies with \zzsun$\la$0.2 are not always consistent with the model predictions for SSFR(\htwo).
Because of the large span of \tdep\ [inverse of SSFR(\htwo)] in our sample,
it could be that the roughly constant timescale in the models is an
oversimplification;
indeed the Krumholz models are not designed for starbursts with potentially short
depletion times and a high-density ISM.
Moreover, there is strong evidence that clumping factors are different
for molecular and atomic phases \citep{leroy13b}, implying that the
use of a single correction factor to compare the predictions by \citet{krumholz11}
with observations is almost certainly an oversimplification.

There could also be other problems with the applicability of the model predictions
to these metal-poor starbursts, since
\aco\ could have additional parameter dependencies beyond metallicity.
The assumption of constant \gnot/$n_H$ may not be valid 
at low metallicities;
the \citet{krumholz11} models assume \gnot\,=\,1,
but the average UV flux \gnot\ at low metallicities is expected to be harder and, 
at a given total stellar mass, more intense because of the warmer stellar 
temperatures. 
There could also be some problem with the assumption that dust opacities
and the capacity of the gas to self-shield from UV radiation vary linearly with metallicity.
As mentioned in Sect. \ref{sec:sample},
even at a similar low metallicity, 
galaxies can have dramatically different dust-to-gas and
dust-to-stars mass ratios \citep{hunt14}.
\citet{schneider15} suggest that this difference is due to the differences in
the density of the cool gas in the ISM, because of the increased efficiency
of grain formation in high-density environments. 
Thus, in addition to metallicity, ISM volume density could also be an important parameter 
for governing both the \hi-to-\htwo\ and the \htwo-to-CO transitions in galaxies.
However, the incidence of high ISM densities at low metallicities is not yet
well established.
More observations of the molecular ISM are needed for metal-poor galaxies,
especially at \zzsun$\la0.1-0.2$ where the validity of CO as a tracer of \htwo\
begins to be called into question.

\section{Summary and conclusions}
\label{sec:conclusions}

We have detected \coone\ with IRAM 30-m observations of a sample of 8 metal-poor dwarf galaxies, 
including a tentative detection at a metallicity of \logoh\,=\,7.7, $\sim$0.1\,\zzsun.
By calculating stellar masses and SFRs for our sample, and comparing
our observations with a  compilation of data from the literature,
we have extended the trend for molecular \tdep\ with stellar specific SFR
by \citet{saintonge11b} to lower metallicities, lower stellar masses,
and higher sSFRs.
In agreement with \citet{saintonge11b} and \citet{huang14}, we conclude that \tdep\
is not constant in galaxies; in our combined sample \tdep\ varies by a factor of
200 or more (from $\la$50\,Myr to $\sim$10\,Gyr) over a range of
$10^3$ in stellar sSFR and \mstar.

By exploiting the correlation of \tdep\ with sSFR, we were able to
constrain \aco\ as a function of metallicity, and found a dependence
of \aco$\,\propto\,$\zzsun$^{-1.9}$, similar to that found by
work based on dust-continuum measurements compared with gas mass
\citep{leroy11,bolatto11}.
In conjunction with atomic gas measurements taken from the literature,
we used our formulation for \aco\ to infer total gas masses, 
and total baryonic masses.
The gas mass relative to the total baryonic mass depends on 
\mstar, stellar sSFR, and O/H; we found significant correlations
for all three parameters in the sense that gas-mass fractions are
higher for galaxies of lower stellar mass, lower metallicity and higher sSFR.
However, despite large scatter, the \hi\ SFR depletion time
is roughly constant for our combined sample, $\sim$3.4\,Gyr
\citep[see also][]{schiminovich10}.

We have examined our data in the context of the star-formation models
by \citet{krumholz11}, and compared the power-law formulation of \aco\
with the theoretically motivated exponential formulation.
The predicted ratios of SFR and gas mass compare extremely well for the CO data, 
but the SSFR(\htwo) inferred from either form of \aco\ are inconsistent for \zzsun$\la$0.2.
A residual trend of SSFR(\htwo) remains even when it should be
invariant with metallicity according to the models.
The cause of the disagreement is not clear 
but could arise from the starburst nature of our metal-poor sample;
more observations of molecular tracers at low metallicity are needed to 
perform a more extended comparison with model predictions.

%%%%%%%%%%%%%%%%%%%%%%%% acknowledgments
\begin{acknowledgements}
We are grateful to the IRAM staff, both in Granada
and at Pico Veleta, for ably managing the logistics and the telescope/receiver operations.
The referee's comments were timely and insightful, and improved the manuscript.
We thank the International Space Science Institute (Bern) for hospitality during the 
conception of this paper, and
Amelie Saintonge and Mei-Ling Huang who kindly gave us their COLDGASS data
in digital form.
LKH acknowledges support from PRIN-INAF 2012/13.
We made use of the NASA/IPAC Extragalactic Database (NED).
\end{acknowledgements}
%%%%%%%%%%%%%%%%%%%%%%%%%%%%%%%%%%%%%

\end{document}